\RequirePackage{fix-cm}
\documentclass[smallcondensed]{svjour3}     
\smartqed  
\usepackage{graphics}
\usepackage{graphicx}
\usepackage{float}
\usepackage[justification=centering]{caption}
\usepackage{amssymb}
\usepackage{amsmath}
\usepackage{amssymb}
\usepackage{amssymb}
\usepackage{amsmath}
\usepackage{amssymb}
\usepackage{color,soul}
\usepackage{float}
\usepackage[justification=centering]{caption}
\usepackage{subfig}
\usepackage{setspace}
\usepackage{relsize}
\usepackage[margin=1in]{geometry}
\usepackage{lineno}
\usepackage{hyperref}
\usepackage{setspace}
\usepackage{mathrsfs}
\usepackage{kpfonts}
\usepackage{mathtools}
\usepackage{float}
\usepackage{amsmath}
\usepackage{esvect}
\usepackage{epstopdf}
\usepackage{placeins}
\usepackage{cite}
\journalname{}
\newcommand{\bd}{\begin{definition}}
    \newcommand{\ed}{\end{definition}}
\newcommand{\br}{\begin{remark}}
    \newcommand{\er}{\end{remark}}
\newcommand{\nn}{\nonumber}
\newcommand{\bea}{\begin{eqnarray}}
\newcommand{\eea}{\end{eqnarray}}
\newtheorem{thm}{Theorem}
\newcommand{\beann}{\begin{eqnarray*}}
    \newcommand{\eeann}{\end{eqnarray*}}

\begin{document}

\title{Vertical sediment concentration distribution revisited with shear-induced diffusivity: An explicit series solution based on homotopy analysis method
}

\titlerunning{Vertical sediment concentration distribution revisited...}        

\author{Punit Jain \and Manotosh Kumbhakar \and
        Koeli Ghoshal 
}


\institute{Punit Jain\\
              Department of Mathematics, Indian Institute of Technology Kharagpur, Kharagpur 721302, India \\
              \email{punitjain51@gmail.com}\\\\
              Manotosh Kumbhakar\\
              Department of Biological and Agricultural Engineering, Texas A\&M University, College Station, TX 77843, USA\\
              \email{manotosh.kumbhakar@exchange.tamu.edu}\\\\
              Koeli Ghoshal\\
              Department of Mathematics, Indian Institute of Technology Kharagpur, Kharagpur 721302, India \\
              \email{koeli@maths.iitkgp.ac.in}
}

\date{Received: date / Accepted: date}

\maketitle
\begin{abstract}
The present study revisits the vertical distribution of suspended sediment concentration in an open channel flow with a special attention to sediment diffusion coefficient. If turbulent diffusivity is considered to follow a parabolic-type profile, the diffusivity coefficient is zero at the bed and very small near the bed; so alone it may not be enough to diffuse the particles from bed-load layer to suspension region. Leighton \& Acrivos (J. Fluid Mech., vol. 181, 1987, pp. 415-439) introduced the idea of  shear-induced diffusion that arises due to the hydrodynamic interactions between solid particles. This work considers the Hunt diffusion equation incorporating the concept of shear-induced diffusion and reinvestigates the vertical sediment concentration profile. Analytical solution is derived using a non-perturbation approach, namely Homotopy Analysis Method (HAM), and is verified with numerical solution as well as compared with available experimental data. The behaviour of the shear-induced diffusion coefficient with vertical distance and varying particle diameters have been interpreted physically. In addition, the effects of important turbulent factors such as inverse of Schmidt number, hindered settling velocity on concentration profile, have been investigated considering relevant sets of experimental data.
\end{abstract}
\section{Introduction}\label{introduction}
Transport of sediment in open channel flow has long  been an important area of research in fluvial hydraulics \cite{rouse1937modern,hunt1954turbulent,zagustin1968sediment,van1984sediment,wang1992theoretical,mazumder2006velocity,ng2008dispersion,kundu2014effects,ghoshal2014analytical,pal2016lag,kumbhakar2017renyi} and has wide application in industry and geophysical research. Better understanding of sediment transport mechanism in natural watercourses needs knowledge of open channel flow. Sediment movement in an open channel flow is generally classified into two categories : (1) Bed-load and (2) Suspended load \cite{julien2010erosion}. Very near to the bed, there is a thin layer known as bed-load layer where the particles move as bed-load showing three types of movement- sliding, rolling and saltating whereas in suspended load, the sediment particles are carried out by the main flow remaining in suspension. In the study of suspended load, researchers' main interest is to find the spatial distribution of sediment concentration as it helps in determining the suspended load transport rate, bed-load transport rate and many others. In the present model, our main focus is on the vertical distribution of suspended sediment concentration in open channel turbulent flow.
\par
The governing equation for vertical concentration distribution of sediments, also termed as one-dimensional advection-diffusion equation, originates from Fick's law of diffusion. Rouse \cite{rouse1937modern} was the pioneer in this field to express the vertical concentration analytically as a function of flow depth though it can not be successfully applied for high concentrated flow. Also that, it shows limitations in predicting concentration near the channel bed and near the water surface ; the reason might be the use of Prandtl-von Karman velocity profile which models velocity in clear water flow and not in sediment laden flow. Hunt \cite{hunt1954turbulent} proposed another analytical model of concentration by  treating the fluid phase and solid phase separately which can be applied for highly concentrated  flow also. Following the works of \cite{rouse1937modern} and \cite{hunt1954turbulent}, numerous investigations have been carried out on vertical concentration distribution in the last few decades. Greimann and Holly Jr \cite{greimann2001two}, Jian et al \cite{jiang2004two}, Zhong et al \cite{zhong2011transport} developed models on the basis of two phase flow approach. Out of many other studies, some \cite{mclean1992calculation,wright2004flow} included effect of stratification, some \cite{kundu2014effects,ghoshal2013influence} included effect of secondary current and some \cite{umeyama1999velocity,kovacs1998prandtl} included effect of concentration in the mixing length. Recently Pal and Ghoshal \cite{pal2017hydrodynamic} developed a model on suspended sediment distribution in an open channel turbulent flow by taking into account different prominent hydrodynamics mechanisms and compared their model with a broad range of previously published models. Though wide research has been carried out to study vertical concentration in a turbulent flow, but still no model can be claimed to be the best one due to highly unpredictable behavior of turbulence which will never allow any model to be applicable for all types of data under all flow conditions. So the topic of studying concentration  in a turbulent flow will ever be under research.
\par
Turbulent diffusivity which plays an important role in modelling concentration, has been found to have three types of profiles: constant, linear and parabolic-type \cite{rijn1984sediment,graf2002suspension}. Amongst these three, under steady-uniform and unidirectional open channel flows, it is found that the parabolic-type profile estimates experimental data better than the others \cite{cellino1999sediment,graf2002suspension} This profile follows from the well-known logarithmic profile for stream-wise velocity and a linear profile for Reynolds shear stress. The present study considers the parabolic-type profile for turbulent diffusion coefficient. But this creates a serious problem in modeling of vertical concentration because with this assumption, the diffusivity coefficient becomes zero at the bed. So the turbulent diffusion coefficient is generally calculated at a certain distance from the bed where the diffusion coefficient is large enough to lift the  bed particles in suspension. Now the question naturally arises : in between the bed-load layer and the small distance from the bed-load layer where the diffusion coefficient is very small, who is responsible to diffuse the particles from bed-load layer to suspension? So there must be some other method of estimating the diffusivity of solid particles very near to the bed. Whenever a fluid is in motion, shear-induced diffusion force occurs due to fluid shear stress that can be accounted in both laminar and turbulent flows \cite{tiwari2009modeling}. Shear-induced diffusion (SID) is a phenomenon in which particles in suspension exhibit diffusive motion during flow due to the interactions between them. This diffusion force has similar behavior as turbulent dispersion force despite the physical sources for these two forces are different \cite{tiwari2009modeling}. So the present study considers that the diffusion of particles from bed-load layer to a small height in suspension from bed-load layer occurs due to sediment diffusion together with shear-induced diffusion and aims to develop a solid particle transportation model by breaking the total diffusivity coefficient into two parts - sediment diffusivity coefficient and shear-induced diffusivity coefficient following Ramadan et al. \cite{ramadan2001application}. In addition,  the diffusivity of solid particles is taken to be different from diffusivity of the fluid as suggested by many researchers \cite{cellino1999sediment,graf2002suspension,van1984sediment}. Also that, the study takes into account the continuity equation of solid and liquid phase separately and the hindered settling effect on a particle due to presence of other particles.
\par
\begin{figure}[!htb]
\centering
\includegraphics[scale=.8]{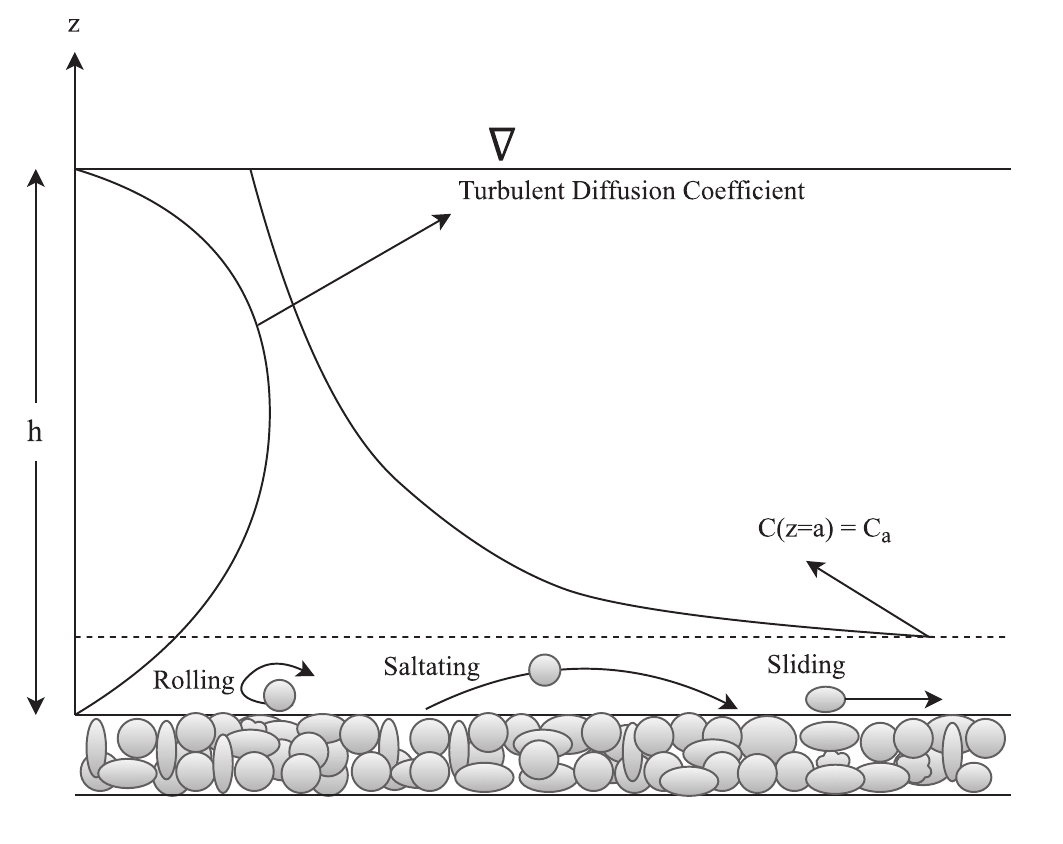}
\caption{Schematic diagram of sediment transport process in open channel flow.} \label{1}
\end{figure}

All the previously mentioned works, except Rouse \cite{rouse1937modern} and Hunt \cite{hunt1954turbulent}, provided the solutions of their models numerically. In comparison to numerical models, the number of analytical models are very few in literature \cite{ghoshal2013influence,absi2010concentration}. It is due to the fact that the inclusion of several turbulent mechanisms makes the governing equation highly non-linear and hence analytical solution may need special mathematical tool. But numerical models result in discrete points of a curve and often time consuming. Apart from that, numerical results does not provide whole understanding of a non-linear problem. Honestly, numerical and analytical methods of non-linear problems have their own advantages and disadvantages and it is not wise to do only one neglecting the other. So the present study also aims to provide an explicit analytical (series) solution of the developed model.   Out of the many analytical methods to solve a non-linear differential equation, the present work adopts Homotopy Analysis Method which is based on homotopy of topology. Traditionally, for obtaining analytical approximation of non-linear problems in science and engineering, perturbation and asymptotic techniques were broadly used. But these techniques are dependent on small or large parameters and mostly applicable for weak non-linear problems only. Liao \cite{liao1992proposed} showed that HAM is not only independent on small/large parameters but also assures the convergence of series solution through a convergence control parameter. Since then, HAM has been successfully applied to solve many non-linear problems \cite{liao1999explicit,inc2008application,yao2009series,vajravelu2011convective,qian2014homotopy,patel2016homotopy,singh2017solution,renuka2019entropy} , though applicability of HAM in the area of sediment transport is too limited. To the best of the authors' knowledge, very few works \cite{kumbhakar2018ham,kundu2019analytical,mohan2019semianalytical,jain2020explicit} showed the application of HAM in the area of sediment transport. The present study finds an explicit series solution through HAM by solving Hunt \cite{hunt1954turbulent} equation which has been modified by using total diffusion coefficient as a sum of sediment diffusion coefficient, shear-induced diffusivity coefficient against concentration gradient and shear-induced diffusivity coefficient against the shear gradient together with the hindered settling effect and a non-unit ratio of sediment and momentum diffusion coefficient. The analytical approximate solution is compared with numerical solution and the effect of shear-induced diffusion coefficient is observed near the channel bed. Finally, the model has been validated through laboratory channel data available in literature. \ref{1}
\section{Formulation of mathematical model}\label{S_mathematical model}
\subsection{Generalized advection diffusion Equation}\label{S_Advection}
The generalized three-dimensional advection-diffusion equation in incompressible fluid flow is obtained as \cite{julien2010erosion}:
\bea
&&\hspace{-1cm}\frac{\partial C}{\partial t}+u\frac{\partial C}{\partial x}+v\frac{\partial C}{\partial y}+w\frac{\partial C}{\partial z}+C\bigg(\frac{\partial u}{\partial x}+\frac{\partial v}{\partial y}+\frac{\partial w}{\partial z}\bigg) \nn \\
&&\hspace{1cm}=\frac{\partial}{\partial x}\bigg[(\varepsilon_{m}+\varepsilon_{sx})\frac{\partial C}{\partial x}\bigg]+\frac{\partial}{\partial y}\bigg[(\varepsilon_{m}+\varepsilon_{sy})\frac{\partial C}{\partial y}\bigg]
+\frac{\partial}{\partial z}\bigg[(\varepsilon_{m}+\varepsilon_{sz})\frac{\partial C}{\partial z}\bigg]\label{advection diff}
\eea
where $C$ is the volumetric suspended sediment concentration (dimensionless), $x$, $y$ and $z$ represent streamwise, transverse and vertical directions respectively, and $u$, $v$ and $w$ are the time-averaged velocity components in $x$, $y$ and $z$-direction respectively. The molecular diffusivity is $\varepsilon_{m}$ and $\varepsilon_{sx}$, $\varepsilon_{sy}$ and $\varepsilon_{sz}$ are the sediment diffusivities in $x$, $y$ and $z$-direction respectively. In viscous flow, the molecular diffusion is dominant ($\varepsilon_{m}\neq 0$) and the turbulent diffusion does not exist. On the other hand, in a turbulent flow, molecular diffusion is negligible ($\varepsilon_{m}\approx 0$) in comparison to the turbulent diffusion. 
In case of steady, uniform flow where sediment concentration varies only in the vertical direction, Eq (\ref{advection diff}) is simplified to
\bea
w\frac{\partial C}{\partial z}=\frac{\partial}{\partial z}\bigg(\varepsilon_{sz}\frac{\partial C}{\partial
z}\bigg) \label{advection 1}
\eea
In Eq. (\ref{advection 1}), the vertical velocity component $w$ is replaced by the settling velocity $-w_{s}$ (negative sign is due to downwards motion) of sediment and the solid or sediment diffusivity in $z$-direction $\varepsilon_{sz}$ is denoted as $\varepsilon_{s}$. Integration of Eq. (\ref{advection 1}) yields
\bea
\varepsilon_{s}\frac{\partial C}{\partial z}+C w_{s}=0 \label{rouse}
\eea
Eq. (\ref{rouse}) physically represents that the equilibrium of sediment suspension occurs by balancing the upward sediment flux due to turbulence and the downward settling due to gravity. The sediment diffusion coefficient $\varepsilon_{s}$ is related to the turbulent diffusion coefficient $\varepsilon_{t}$ by the relation \cite{majumdar1967diffusion} 
\begin{eqnarray}
\varepsilon_{s}=\beta \varepsilon_{t} \label{Eq.2}
\end{eqnarray}
Rouse \cite{rouse1937modern} was the first to derive an analytical solution for Eq. (\ref{rouse}) using a parabolic-type profile for $\varepsilon_{t}$ which is given as :
\bea
\varepsilon_{t}=\kappa u_{\ast}z\big(1-\frac{z}{h}\big) \label{Eq.3}
\eea
The solution obtained by him is known as the Rouse equation which reads as follows
\begin{eqnarray}
\frac{C}{C_{a}}&=&\left(\frac{1-\hat{z}}{\hat{z}}
\frac{\hat{a}}{1-\hat{a}} \right)^{\frac{w_{s}}{\kappa \beta u_{\ast}}}\label{rouse 1}
\end{eqnarray}
where $C_{a}$ is the reference concentration at the dimensionless reference level $\hat a~\Big(=\frac{a}{h}\Big)$ with $a$ as the reference level and $\kappa$  the von Karman as the constant ($=0.41$). Here $u_{\ast}$ is the shear velocity and
$\hat{z}~\Big(=\frac{z}{h}\Big)$ is the normalized vertical height where $z$ is vertical distance from channel bed.
Further modification was done by Hunt \cite{hunt1954turbulent} who considered the state of equilibrium of solid (suspended sediment) and fluid phases, separately. The equation of solid phase is obtained as:
\bea
-w_{b}\frac{\partial C}{\partial z}-C \frac{\partial w_{b}}{\partial z}+\frac{\partial}{\partial z}\bigg(\varepsilon_{sz}\frac{\partial C}{\partial z}\bigg)=0 \label{h 1}
\eea
where $w_{b}$ is the velocity of suspended solid particle in $z$-direction. Likewise, for the fluid phase, the equation is given by
\bea
-w\frac{\partial C}{\partial z}+(1-C) \frac{\partial w}{\partial z}+\frac{\partial}{\partial z}\bigg(\varepsilon_{tz}\frac{\partial C}{\partial z}\bigg)=0 \label{h 2}
\eea
The time-averaged vertical velocity component $w_{b}$ of sediment particles is equal to the sum of the vertical velocity component $w$ of fluid together with the settling velocity of the sediment particles in clear water $-w_{s}$ i.e.
\bea
w_{b}=w-w_{s}\label{settling 1}
\eea
One can eliminate $w_{b}$  and $w$ from the Eqs. (\ref{h 1}) and (\ref{h 2}) by using Eq. (\ref{settling 1}) and arrive at the equation
\bea
\varepsilon_{s}\frac{dC}{dz}+C(\varepsilon_{t}-\varepsilon_{s})\frac{dC}{dz}+C(1-C)w_{s}=0\label{h 3}
\eea
Hunt \cite{hunt1954turbulent} assumed the solid and turbulent diffusivities to be same, $i.e.$ $\varepsilon_{t}=\varepsilon_{s}$ to simplify the solution. Eq. (\ref{h 3}) is widely used in suspension study to find distribution of sediment along a vertical in an open channel flow over an erodible sediment bed. The primary objective of the present study is to modify Eq. (\ref{h 3}) by incorportaing the effect of shear-induced diffusion which is described in the following section.
\subsection{Modification of governing equation with shear induced diffusion}\label{S_modification hunt}
It has  already been  discussed  that diffusivity coefficient may not be responsible alone to diffuse the particle from bed-load layer to suspension since the diffusion coefficient is very small near the bed-load layer and becomes zero at the bed. The reason behind the diffusion of particles in between the bed-load layer and a short distance from it might be due to the effect of some other factors. Leighton and Acrivos \cite{leighton1987measurement} reported that in concentrated sheared suspensions, particles flux is induced from regions of high shear to low and from high concentration to low. This phenomenon is called the shear-induced diffusion that arises from the hydrodynamic interaction among the particles \cite{zhang1994viscous}. Zhang and Acrivos \cite{zhang1994viscous} provided a theoretical analysis of horizontal circular pipe flow where the solid particles occupying the bottom portion of the pipe were initially stratified. For channel flow, total shear-induced diffusion coefficient $\varepsilon_{scg}$ can be expressed as \cite{zhang1994viscous,ramadan2001application}:
\bea
\varepsilon_{scg}=-\varepsilon_{cg}-D_{s}\frac{\displaystyle{\frac{d^{2}u}{dz^{2}}}}{\displaystyle{\frac{dC}{dz}}} \label{Diffusion Coefficient}
\eea
where the first term $\varepsilon_{cg}$ in the right side is the shear-induced diffusion coefficient against the concentration gradient and the second term $D_{s}\frac{d^{2}u/dz^{2}}{dC/dz}$ is an equivalent shear-induced diffusion coefficient against the shear gradient which is denoted by $\varepsilon_{sg}$ . So in this study, we replace sediment diffusion coefficient in Eq. (\ref{h 3}) with the total diffusion coefficient $(\varepsilon)$ which is the sum of  sediment diffusion coefficient ($\varepsilon_{s}$ ), shear-induced diffusivity coefficient against concentration gradient $(\varepsilon_{cg})$ and shear-induced diffusivity coefficient against the shear gradient $(\varepsilon_{sg})$ \cite{ramadan2001application}. With all the above assumptions, Eq. (\ref{h 3}) is written as
\bea
\varepsilon\frac{dC}{dz}+C(\varepsilon_{t}-\varepsilon)\frac{dC}{dz}+C(1-C)w_{s}=0\label{hunt}
\eea
where the total diffusivity coefficient $\varepsilon$ is given as \cite{ramadan2001application} :
\bea
\varepsilon&=&\varepsilon_{s}+\varepsilon_{cg}+\varepsilon_{sg} \label{total diff}
\eea
The present study considers Eq. (\ref{hunt}) as the governing equation for distribution of concentration of sediment in an open channel flow. To solve Eq (\ref{hunt}), one should know  each of sediment diffusion coefficient $(\varepsilon_{s})$, shear-induced diffusivity coefficient against concentration gradient $(\varepsilon_{cg})$ and shear-induced diffusivity coefficient against the shear gradient $(\varepsilon_{sg})$ present in the total diffusion coefficient. About the sediment diffusion coefficient, it has already been discussed (see Eq. (\ref{Eq.2}) and (\ref{Eq.3})). The other part of total diffusion coefficient is shear-induced diffusion coefficient  which is the sum of shear-induced diffusion coefficient against shear gradient and shear-induced  diffusion  coefficient against concentration gradient. Expression for $\varepsilon_{cg}$ was revealed by  \cite{zhang1994viscous,ramadan2001application} as :
\bea
\varepsilon_{cg}&=&\frac{D_{p}^{2}}{4}\frac{du}{dz}\Big(0.43C+0.65C^{2}\frac{1}{k_{\nu}}\frac{dk_{\nu}}{dC}\Big) \label{CG1}
\eea
where $D_{p}$ is used to denote the particle diameter, $u$ is the local flow velocity which is given by Prandtl \cite{prandtl1933recent} as follows :
\bea
u=\frac{u_{\ast}}{\kappa}ln\bigg(\frac{z}{z_{0}}\bigg)\label{velocity}
\eea
here $z_{0}$ is the distance from the bed where the velocity is hypothetically equal to zero, and symbol $k_{\nu}$ stands for relative suspension viscosity, which is the ratio of suspension viscosity to pure fluid viscosity and was estimated using a relation developed as \cite{krieger1972rheology}:
\bea
k_{\nu}&=&\bigg(1-\frac{C}{c_{m}}\bigg)^{-1.82} \label{KV}
\eea
where $c_{m}$ is a solid maximum packing fraction, which is calculated by previous study \cite{cheng1999analysis} to be 0.5812. Therefore, by using Eqs. (\ref{velocity}) and (\ref{KV}) in Eq. (\ref{CG1}), we can write the simplified form for the diffusion coefficient against the concentration gradient as :
\bea
\varepsilon_{cg}&=&\frac{D_{p}^{2}}{4}\frac{u_{\ast}}{ \kappa z}\Bigg\{0.43C+2.04C^{2}\Big(1-\frac{C}{0.5812}\Big)^{-1}\Bigg\} \label{CG2}
\eea
Similarly, an expression for diffusion coefficient against the shear gradient \cite{ramadan2001application} can be written as:
\bea
\varepsilon_{sg}&=&0.43\frac{C^{2} D_{p}^{2}}{4} ~\frac{ \displaystyle{ \frac{d^{2}u}{dz^2}}}{ \displaystyle{\frac{dC}{dz}}}\label{SG}
\eea
Apart from diffusion coefficients, the solution of Eq. (\ref{hunt}) needs an expression of settling velocity $w_s$ which is discussed in subsubsection \ref{S_Settling Velocity}. As mentioned previously, the present work considers non unit ratio of sediment and momentum diffusion coefficients and hence an expression of $\beta=\frac{\varepsilon_{s}}{\varepsilon_{m}}$ is required which is discussed in subsubsection \ref{S_beta}.
\subsubsection{Settling velocity}\label{S_Settling Velocity}
A proper assessment of the settling velocity of sediment particles is fundamental to the modelling of sediment suspension. The settling velocity of sediment particles in a sediment-laden flow is less than settling velocity in clear fluid, which is commonly known as hindered settling. Richardson and Zaki \cite{richardson1954sedimentation} proposed the expression for $w_{s}$ which is given as :
\bea
w_{s}&=&w_{0}(1-C)^{\eta} \label{settling}
\eea
where $w_0$ is the settling velocity of particle in clear water and $\eta$ is the exponent of reduction of settling velocity that depends on the particle Reynolds number $Re$ as follows :
\bea
\eta=\begin{cases}4.65 \quad& \mbox{when }Re<0.2\\{}4.4Re^{-0.03} \quad& \mbox{when }0.2<Re<1\\{}4.4Re^{-0.1} \quad& \mbox{when } 1<Re<500 \\{}2.4 \quad& \mbox{when }Re>500   \end{cases}
\eea
where $Re=\frac{w_{0}D_{p}}{\nu_{f}}$, $D_{p}$ is the particle diameter and $\nu_f$ denotes the kinematic viscosity of clear fluid. Here settling velocity of particle in clear fluid is calculated from the widely used formula given as \cite{cheng1997simplified}
\bea
w_{0}=\frac{\nu_{f}}{D_p}\left(\sqrt{25+1.2D_{\ast}^{2}}-5\right)^{1.5}
\eea
where $D_{\ast}\Bigg(=\Big(\frac{\vartriangle g}{\nu_{f}^{2}}\Big)^{(\frac{1}{3})}D_{p}\Bigg)$ is the dimensionless particle diameter, $g$ is the gravitational acceleration and $\vartriangle=s-1$, $s$ being the specific gravity of sediment particle.
\subsubsection{Non-equality of inverse of Schmidt number}\label{S_beta}
In the study of vertical concentration distribution, $\beta$ $(=\frac{\varepsilon_{s}}{\varepsilon_{t}})$, known as inverse of Schmidt number, plays an important role. A large number of different experiments reported the non-equality of $\varepsilon_{s}$ and $\varepsilon_{t}$ for different flow conditions. Cellino and Graf \cite{cellino1999sediment} determined that $\beta$ is less than 1 for open channel flow under both capacity and non-capacity conditions. Graf and Cellino \cite{graf2002suspension} experimentally determined $\beta<1$ without bedform and $\beta>1$ with bedform over an erodible sediment bed. However, experimental results revealed that $\beta\approx1$ for finer sediment particles only, and $\beta<1$ for coarse sediment particles \cite{majumdar1967diffusion,brush1962study,jobson1970vertical}. Thus, the assumption $\beta\approx1$ may not be physically reasonable; so in our model, we assume that $\beta\neq 1$.
\par
Many researchers \cite[]{graf2002suspension,van1984sediment,wren2005distributions,jain2018mathematical} reported that $\beta$ is a function of normalized settling velocity $\frac{w_{0}}{u_{\ast}}$. Pal and Ghoshal \cite{pal2016effect} established two relations of $\beta$ for dilute and non-dilute by fitting the Rouse equation with a large number of data sets of different researchers. Their study claims that $\beta$ values do not depend only upon normalized settling velocity, but depends also on reference level and reference concentration. They gave the following
expressions as:
\bea
\beta=\begin{cases}0.033\Big(\frac{w_{0}}{u_{\ast}}\Big)^{0.931}C_{\hat{a}}^{-0.118}\hat{a}^{-1.196}\quad& \mbox{for dilute flow} \\2.2040\Big(\frac{w_{0}}{u_{\ast}}\Big)^{0.667}C_{\hat{a}}^{0.017}\hat{a}^{0.178} & \mbox{for non-dilute flow}  \end{cases} \label{beta}
\eea
In this model, we will use the expression of $\beta$ given by Eq. (\ref{beta}).
\par\par
Now, after using Eqs. (\ref{total diff}),(\ref{CG2}), (\ref{SG}) and  (\ref{settling}), Eq. (\ref{hunt}) is written as:
\bea
&&\hspace{-1.5cm}\Bigg\{\kappa\beta u_{\ast} z\Big(1-\frac{z}{h}\Big)+\frac{D_{p}^{2}}{4}\frac{u_{\ast}}{\kappa z}\Bigg(0.43C+2.04C^{2}\Big(1-\frac{C}{0.5812}\Big)^{-1}\Bigg)-0.43
\frac{\displaystyle{ \frac{  u_{\ast}C^2 D_{p}^{2}}{4 \kappa z^{2}}}}{\displaystyle{\frac{dC}{dz}}} \Bigg\} (1-C)\frac{dC}{dz}\nn\\
&&\hspace{3cm}+\kappa u_{\ast} z\Big(1-\frac{z}{h}\Big)C\frac{dC}{dz}+w_{0}C(1-C)^{\eta+1}=0\label{hunt1}
\eea
Before proceeding further, we rearrange Eq. (\ref{hunt1}) in non-dimensional form and using the series expansion for $\Big(1-\frac{C}{0.5812}\Big)^{-1}$ as $\frac{C}{0.5812}<1$
and retaining terms up to second  order, Eq. (\ref{hunt1}) changes to
\bea
&&\hspace{-1.0cm}\Bigg[\beta\kappa\hat{z}(1-\hat{z})+\frac{\hat{D}_{p}^2}{4\kappa\hat{z}}\bigg\{0.43C+2.04C^2\Big(1+\frac{C}{0.5812}+\Big(\frac{C}{0.5812}\Big)^{2}\Big)\bigg\} \nn \\
&&\hspace{1cm}-0.43 \frac{\displaystyle{\frac{C^{2}\hat{D}_{p}^{2}}{4\kappa \hat{z}^{2}}} }{ \displaystyle{\frac{dC}{d\hat{z}}}} \Bigg](1-C)\frac{dC}{d\hat{z}}
+\kappa\hat{z}(1-\hat{z})C\frac{dC}{d\hat{z}}+\hat{w}_{0}C(1-C)^{\eta+1}=0\label{hunt2}
\eea
where $\hat{D}_{p}=\frac{D_{p}}{h}$ is the normalized diameter of particle and $\hat{w}_{0}=\frac{w_{0}}{u_{\ast}}$ is the normalized settling velocity.
Eq. (\ref{hunt2}) can further be written as:
\bea
&&\hspace{-1.3cm}\beta A\frac{dC}{d\hat{z}}+\Big(-\beta A + A + 0.43\frac{B}{\hat{z}}\Big)C\frac{dC}{d\hat{z}}+\frac{B}{\hat{z}}\Bigg(1.605C^2\frac{dC}{d\hat{z}} +2.107C^3\frac{dC}{d\hat{z}}+4.2898C^4\frac{dC}{d\hat{z}}\nn\\
&&\hspace{2cm}-8.432C^5\frac{dC}{d\hat{z}}\Bigg) -0.43 \frac{B}{\hat{z}^2}C^2(1-C)+\hat{w}_{0}C(1-C)^{\eta+1}=0 \label{hunt3}
\eea
where $A=\kappa\hat{z}(1-\hat{z})$ and $B=\frac{\hat{D_{p}}^2}{4\kappa}$.
Equation (\ref{hunt3}) is a first-order highly non-linear ordinary differential equation which represents vertical distribution of suspended sediment concentration.
\section{Series solution by homotopy analysis method (HAM)}\label{S_HAM}
The second objective of the present study is to derive an explicit analytical approximate solution of the generalized diffusion equation (\ref{hunt3}). In order to do this, as mentioned previously, we choose an analytical approximate solution method called Homotopy analysis method developed by Liao \cite{liao1992proposed} which is widely used to solve non-linear differential equations in various fields of science and engineering. To understand this, let us consider the differential Eq. (\ref{hunt3}) as follows :
\bea
\mathscr{N}[C(\hat{z})]&=&0 \label{ham1}
\eea
subject to the boundary condition
\bea
C(\hat{z}=\hat{a})=C_{\hat{a}}
\eea
where $\mathscr{N}$ is a non-linear operator and $\hat{z}$ denotes the independent variable. Liao \cite{liao1992proposed} used the concept of homotopy in topology to construct a one-parameter family of equations in the embedding parameter $q\in[0,1]$, called the zeroth-order deformation equation
\bea
(1-q)\big(\mathscr{L}[\Phi(\hat{z};q)]-\mathscr{L}[C_{0}(\hat{z})]\big)+q\hbar H(\hat{z})\mathscr{N}[\Phi(\hat{z};q)]=0\label{ham2}
\eea
subject to the boundary condition
\bea
\Phi(\hat{a};q)=C_{\hat{a}}
\eea
where $C_{0}(\hat{z})$ is an initial approximation of $C(\hat{z})$ and $\mathscr{L}$ stands for an auxiliary linear operator with the property
\bea
\mathscr{L}f&=&0~~\mbox{when}~~f=0, \label{ham3}
\eea
$\hbar$ is a non-zero auxiliary parameter known as convergence control parameter and $H(\hat{z})$ is a non-zero auxiliary function.
The underlying idea behind HAM is that a continuous mapping is described to relate the solution $C(\hat{z})$ and the unknown function $\Phi(\hat{z};q)$, with the aid of the embedding-parameter $q$. As $q$ increases from $0$ to $1$, $\Phi(\hat{z};q)$ varies from the initial approximation
$C_{0}(\hat{z})$ to the exact solution $C(\hat{z})$ of the non-linear problem. We now define:
\bea
C_{m}(\hat{z})=\frac{1}{m!}\frac{\partial^{m}\Phi(\hat{z};q)}{\partial q^{m}}\biggm|_{q=0} \label{ham4}
\eea
where $C_{m}(\hat{z})$ is called the $m$-th order deformation derivative. As $\Phi(\hat{z};q)$  is also dependent upon the embedding parameter $q\in[0,1]$, one can expand it into the power of $q$ by using the Maclaurin series expansion
\bea
\Phi(\hat{z};q)=C_{0}(\hat{z})+\mathlarger{\sum}_{m=1}^{\infty}C_{m}(\hat{z}) q^{m} \label{ham5}
\eea
According to Liao \cite{liao2003beyond}, convergence of the series totally depends upon $\mathscr{L}$, $ H(\hat{z})$, $C_{0}(\hat{z})$, and $\hbar$. Hence it is assumed that $\mathscr{L}$, $ H(\hat{z})$, $C_{0}(\hat{z})$, and $\hbar$ are properly chosen so that the series Eq. (\ref{ham5}) is convergent at $q = 1$. Then, at $q = 1$, the series becomes:
\bea
C(\hat{z})=C_{0}(\hat{z})+\mathlarger{\sum}_{m=1}^{\infty}C_{m}(\hat{z})\label{ham6}
\eea
The above series is called homotopy-series solution which satisfies the original equation $\mathscr{N}[C(\hat{z})]=0$.
Eq. (\ref{ham6}) provides the relationship between the initial approximation $C_{0}(\hat{z})$ and the exact solution $C(\hat{z})$. However, the higher order approximations $C_{m}(\hat{z})$ for $m\geq1$ are still unknown to us. Liao \cite{liao1992proposed} showed that the higher order terms can be obtained by differentiating the zeroth-order deformation equation Eq. (\ref{ham2}) (with $\mathscr{H}(\Phi(\hat{z};q);q,\hbar, H)=0$) $m$ times with respect to $q$, and setting $q=0$, and finally dividing by $m!$, which can be expressed as:
\bea
\mathscr{L}[C_{m}(\hat{z})-\chi_{m}C_{m-1}(\hat{z})]=\hbar H(\hat{z})R_{m}(\vv{C}_{m-1})
\label{ham7}
\eea
subject to the boundary conditions
\bea
C_{m}(\hat{a})=0,\mbox{for}~m=1,2,3,\cdot\cdot\cdot
\eea
where
\bea
\chi_{m}=\begin{cases}0\quad& \mbox{if~}m\leq1 \\{}1 & \mbox{otherwise }  \end{cases} \label{ham8}
\eea
and
\bea
R_{m}(\vv{C}_{m-1})=\frac{1}{(m-1)!}\frac{\partial^{m-1}\mathscr{N}[\Phi(\hat{z};q)]}{\partial q^{m-1}}\biggm|_{q=0} \label{ham9}
\eea
As mentioned in \cite{liao2003beyond}, in this method, one has great freedom to select auxiliary function , linear operator and parameters in order to ensure the convergence of approximate solutions and to increase both the rate and region of convergence. But it is very difficult to check the convergence region and rate by taking different linear operators, auxiliary functions and  parameters. To overcome this problem, Liao \cite{liao2003beyond} proposed some generalized rule of solution expressions such as rule of coefficient ergodicity and rule of solution existence in order to choose the initial approximation, linear operator and the auxiliary function.  The auxiliary parameter $\hbar$ plays a vital role in controlling and adjusting the convergence rate and region of the series solution obtained through HAM.
\par
For Eq. (\ref{hunt3}), a single-term linear operator is selected based on Liao's rule of solution expression as
\bea
\mathscr{L}[\Phi(\hat{z};q)]=\frac{\partial \Phi(\hat{z};q)}{\partial \hat{z}} \label{ham10}
\eea
The non-linear operator is as follows:
\bea
\mathscr{N}[\Phi(\hat{z};q)]&=&\beta A \frac{d\Phi}{d\hat{z}} + (-\beta A+A+0.43 \frac{B} {\hat{z}})\Phi\frac{d\Phi}{d\hat{z}} +\frac{B}{\hat{z}}\Bigg(1.605\Phi^{2}\frac{d\Phi}{d\hat{z}}+2.107\Phi^{3}\frac{d\Phi}{d\hat{z}}\nn\\
&&\hspace{-1.0cm}+~ 4.2898\Phi^{4}\frac{d\Phi}{d\hat{z}}-8.432\Phi^5\frac{d\Phi}{d\hat{z}}\Bigg)-0.43 \frac{\hat{B}}{\hat{z}^2}\Phi^2(1-\Phi)+\hat{w}_{0}\Phi(1-\Phi)^{\eta+1}
\eea
The auxiliary function $H(\hat{z})$ can be taken as 1 to avoid difficulty in the computation \cite{vajravelu2013nonlinear}. The next step is to find out $R_{m}$'s from Eq. (\ref{ham9}). The closed-form for $R_{m}(\vv{C}_{m-1})$ corresponding to Eq. (\ref{hunt3}) is given as follows:
\bea
R_{m}(\vv{C}_{m-1})&=&\beta A \dot{C}_{m-1} + \Big(-\beta A+A+0.43\frac{B}{\hat{z}}\Big)\mathlarger{\sum}_{i=0}^{m-1}C_{m-i-1}\dot{C}_{i} \nn \\ &+&\frac{B}{\hat{z}}\Bigg(1.605\mathlarger{\sum}_{i=0}^{m-1}\dot{C}_{m-i-1}\sum_{j=0}^{i}C_{j}C_{i-j}+2.107\mathlarger{\sum}_{i=0}^{m-1}\dot{C}_{m-i-1}\sum_{j=0}^{i}C_{i-j}\sum_{k=0}^{j}C_{k}C_{j-k}\nn\\ &+&4.2898\mathlarger{\sum}_{i=0}^{m-1}\dot{C}_{m-i-1}\sum_{j=0}^{i}C_{i-j}\sum_{k=0}^{j}C_{j-k}\sum_{l=0}^{k}C_{l}C_{k-l}\nn\\
&-&8.432\mathlarger{\sum}_{i=0}^{m-1}\dot{C}_{m-i-1}\sum_{j=0}^{i}C_{i-j}\sum_{k=0}^{j}C_{j-k}\sum_{l=0}^{k}C_{k-l}\sum_{n=0}^{l}C_{n}C_{l-n}\Bigg) \nn\\
&-&0.43\frac{B}{\hat{z}}  \Big(\mathlarger{\sum}_{i=0}^{m-1}C_{m-i-1}C_{i}-\mathlarger{\sum}_{i=0}^{m-1}C_{m-i-1}\sum_{j=0}^{i}C_{j}C_{i-j}\Big)+\hat{\mathscr{D}}_{m-1}[f(\Phi)]\bigm|_{q=0}\label{ham12}
\eea
where $\hat{\mathscr{D}}_{m}=\frac{1}{m!}\frac{\partial^{m}}{\partial q^{m}}$ ,  $f(\Phi)=\hat{w}_{0}\Phi(1-\Phi)^{\eta+1}$
and
\bea
\hat{\mathscr{D}}_{m}[f(\Phi)]=\begin{cases}\mathlarger{\sum}_{k=0}^{m-1}\left(1-\frac{k}{m}\right)\hat{\mathscr{D}}_{m-k}(\Phi)\frac{\partial}{\partial \Phi}{\hat{\mathscr{D}}_{k}[f(\Phi)]}\quad& \mbox{if } m\geq 1\\{}f(\Phi) & \mbox{if~}m=0
\end{cases} \label{ham13}
\eea
Following Liao \cite{liao2012homotopy}, the proof of Eq. (\ref{ham13}) is given in \emph{Appendix}. We can find the $m$ th term $C_m$ from Eq. (\ref{ham7}) by applying the inverse of the linear operator ( Eq. (\ref{ham10})) as
\bea
C_m(\hat{z})=\chi_{m}C_{m-1}+\hbar \int_{\hat{a}}^{\hat{z}}R_{m}(\vv{C}_{m-1})d\hat{z} \label{ham14}
\eea
So, the $M$-th order approximated HAM based series solution can be obtained as follows
\bea
C(\hat{z})\approx \sum_{n=0}^{M}C_{n}(\hat{z})\label{ham15}
\eea
It is clear from Eq. (\ref{ham15}) that if the initial approximation $C_{0}(\hat{z})$ is known, the required order of approximation can be found. In order to get the HAM based series solution, the initial approximation $C_{0}(\hat{z})=C_{\hat{a}}$ is selected.

\section{Result and discussion}\label{S_result}
\subsection{Experimental data considered}
In this section, we analyze the validation of the model through experimental data available in the literature. To that purpose, experiential data of  Coleman \cite{coleman1986effects}, Lyn \cite{lyn1988similarity}, Vanoni \cite{vanoni1946transportation} and Einstein and Chien \cite{einstein1955effects}  have been used. A short description on selected data sets is provided below. Details can be found from the respective literature. The experimental conditions of the data sets that are used for validation of the model are reported in tables.
\par
Coleman \cite{coleman1986effects} used a re-circulatory flume $356~mm$ wide and $15~m$ long to study the effect of suspended sediment on the profile of fluid velocity and suspended load by using sediment  concentration profile. A total of $40$ experimental runs was considered. While keeping the other parameters fixed ( water depth $h=1.69~mm$, channel width $b=356~mm$, channel slop $J=0.002$ and shear velocity $u_{\ast}=0.041~ms^{-1}$), Coleman \cite{coleman1986effects} added a fixed amount of sediment into the flow until the capacity condition of suspended load was reached. From the data sets of Coleman \cite{coleman1986effects}, four experimental runs were considered and are summarized in Table \ref{tab:coleman}.
\begin{table}[htbp]
  \centering
  \caption{Summary of the selected experimental data from Coleman \cite{coleman1986effects}}
    \resizebox{\textwidth}{!}
    {\begin{tabular}{lrrrrrrrrrrr}
   \hline
    \multicolumn{1}{l}{Run}  & \multicolumn{1}{l}{$\beta$} & \multicolumn{1}{l}{$\hat{a}$} & \multicolumn{1}{l}{$h~(cm)$} & \multicolumn{1}{l}{$C_a~(\%)$} & \multicolumn{1}{l}{$w_0~(cm/s)$} & \multicolumn{1}{l}{$u_{\ast}~(cm/s)$} & \multicolumn{1}{l}{$D~(cm)$} & \multicolumn{1}{l}{$\hat{w}_0$} & \multicolumn{1}{l}{$Re$} & \multicolumn{1}{l}{$\eta$} & \multicolumn{1}{l}{$D_{\ast}$} \\ \hline
    3 & 0.71 & 0.035 & 17.2  & 0.17  & 0.66 & 4.1   & 0.0105 & 0.16 & 0.70 & 4.45 & 2.66 \\
    4 & 0.67 & 0.035 & 17.1  & 0.28  & 0.66 & 4.1   & 0.0105 & 0.16 & 0.70 & 4.45 & 2.66 \\
    11 & 0.50 & 0.036 & 16.9  & 1.20   & 0.61 & 4.1   & 0.0100  & 0.15 & 0.61 & 4.47 & 2.53 \\
    13 & 0.55 & 0.035 & 17.0    & 1.40   & 0.66 & 4.1   & 0.0105 & 0.16 & 0.70 & 4.45 & 2.66 \\
    \hline
    \end{tabular}}%
  \label{tab:coleman}%
\end{table}%
\par
Similarly the data set of Lyn \cite{lyn1988similarity} is used for the verification of the proposed model. Lyn \cite{lyn1988similarity} performed experiments in a $26.7 ~cm$ wide narrow open channel, for both equilibrium and starved beds. Four experimental runs from the data of Lyn \cite{lyn1988similarity} are used, and a short summary has been provided in Table \ref{tab:lyn}.
\FloatBarrier
\begin{table}[htbp]
  \centering
  \caption{Summary of the selected experimental data from Lyn \cite{lyn1988similarity}}
     \resizebox{\textwidth}{!}
     {\begin{tabular}{lrrrrrrrrrrr}
   \hline
    \multicolumn{1}{l}{Run}  & \multicolumn{1}{l}{$\beta$} & \multicolumn{1}{l}{$\hat{a}$} & \multicolumn{1}{l}{$h~(cm)$} & \multicolumn{1}{l}{$C_a~(\%)$} & \multicolumn{1}{l}{$w_0~(cm/s)$} & \multicolumn{1}{l}{$u_{\ast}~(cm/s)$} & \multicolumn{1}{l}{$D~(cm)$} & \multicolumn{1}{l}{$\hat{w}_0$} & \multicolumn{1}{l}{$Re$} & \multicolumn{1}{l}{$\eta$} & \multicolumn{1}{l}{$D_{\ast}$} \\ \hline
    1957ST-1A & 0.84 & 0.0740 & 5.69  & 0.102 & 1.79 & 3.74  & 0.019 & 0.48 & 3.40 & 3.89 & 4.81 \\
    1957ST-1B & 0.95 & 0.0740 & 5.68  & 0.041 & 1.79 & 3.69  & 0.019 & 0.48 & 3.40 & 3.89 & 4.81 \\
    1957ST-2A & 0.66 & 0.0735 & 5.84  & 0.334 & 1.79 & 4.25  & 0.019 & 0.42 & 3.40 & 3.89 & 4.81 \\
    1957ST-2D & 0.82 & 0.0724 & 5.74  & 0.050  & 1.79 & 4.34  & 0.019 & 0.41 & 3.40 & 3.89 & 4.81 \\
    \hline
    \end{tabular}}%
  \label{tab:lyn}%
\end{table}%
\FloatBarrier
\par
Experimental data of Vanoni \cite{vanoni1946transportation} has been used to verify the proposed model. Vanoni \cite{vanoni1946transportation} used a $845~mm$ wide and $18~m$ long recirculating flume with an artificially sandy rough bed surface, to perform experiments on the transportation of suspended sediment by water. It was observed that the span-wise distribution of particles changed periodically with time. Four experimental runs from the data of Vanoni \cite{vanoni1946transportation} are used and a short summary has been provided in Table \ref{tab:vanoni}.
\FloatBarrier
\begin{table}[htbp]
  \centering
  \caption{Summary of the selected experimental data from Vanoni \cite{vanoni1946transportation}}
    \resizebox{\textwidth}{!}{ \begin{tabular}{lrrrrrrrrrrr}
   \hline
    \multicolumn{1}{l}{Run}  & \multicolumn{1}{l}{$\beta$} & \multicolumn{1}{l}{$\hat{a}$} & \multicolumn{1}{l}{$h~(cm)$} & \multicolumn{1}{l}{$C_a~(\%)$} & \multicolumn{1}{l}{$w_0~(cm/s)$} & \multicolumn{1}{l}{$u_{\ast}~(cm/s)$} & \multicolumn{1}{l}{$D~(cm)$} & \multicolumn{1}{l}{$\hat{w}_0$} & \multicolumn{1}{l}{$Re$} & \multicolumn{1}{l}{$\eta$} & \multicolumn{1}{l}{$D_{\ast}$} \\ \hline
    7  & 0.61 & 0.05 & 15.1  & 0.219 & 1.36  & 6.10   & 0.016 & 0.22 & 2.18 & 4.06 & 4.04 \\
    14 & 0.56 & 0.05 & 7.1  & 0.343 & 1.36  & 6.30   & 0.016 & 0.21 & 2.18 & 4.06 & 4.04 \\
    15 & 0.81 & 0.05 & 9.0   & 0.198 & 1.36  & 4.51  & 0.016 & 0.30 & 2.18 & 4.06 & 4.04 \\
    19 & 0.66 & 0.05 & 7.2   & 0.049 & 0.61  & 2.97  & 0.010  & 0.20 & 0.61 & 4.62 & 2.52 \\
    \hline
    \end{tabular}}%
  \label{tab:vanoni}%
\end{table}%
\FloatBarrier
\par
Further verification of the proposed model has been done by using the data set of Einstein and Chien \cite{einstein1955effects}.  The effect of heavy sedimentation concentration near the bed on velocity and sediment distribution profiles was observed by them. Four experimental runs from the data of Einstein and Chien \cite{einstein1955effects} are used and a short summary has been provided in Table \ref{tab:Einstein}.
\FloatBarrier
\begin{table}[htbp]
  \centering
  \caption{Summary of the selected experimental data from Einstein and Chien \cite{einstein1955effects}}
    \begin{tabular}{lrrrrrrrrrrr}
    \hline
    \multicolumn{1}{l}{Run}  & \multicolumn{1}{l}{$\beta$} & \multicolumn{1}{l}{$\hat{a}$} & \multicolumn{1}{l}{$h~(cm)$} & \multicolumn{1}{l}{$C_a~(\%)$} & \multicolumn{1}{l}{$w_0~(cm/s)$} & \multicolumn{1}{l}{$u_{\ast}~(cm/s)$} & \multicolumn{1}{l}{$D~(cm)$} & \multicolumn{1}{l}{$\hat{w}_0$} & \multicolumn{1}{l}{$Re$} & \multicolumn{1}{l}{$\eta$} & \multicolumn{1}{l}{$D_{\ast}$} \\ \hline
    S1    & 1.30 & 0.04  & 13.8  & 2.189 & 13.51 & 11.47 & 0.130  & 1.18 & 175.67 & 2.62 & 32.89 \\
    S2    & 1.35 & 0.05 & 12.0    & 4.566 & 13.51 & 12.85 & 0.130  & 1.05 & 175.69 & 2.62 & 32.88 \\
    S6    & 1.09 & 0.04 & 14.3  & 1.057 & 10.62 & 11.82 & 0.090 & 0.90 & 99.85 & 2.78 & 23.78 \\
    S12   & 0.51 & 0.03 & 13.2  & 7.721 & 2.96 & 10.09 & 0.027 & 0.29 & 7.90 & 3.57 & 6.83 \\
    \hline
    \end{tabular}%
  \label{tab:Einstein}%
\end{table}%
\FloatBarrier
\subsection{Verification of the HAM-based solution}
\begin{figure}[!htb]
\centering
\includegraphics[scale=.5]{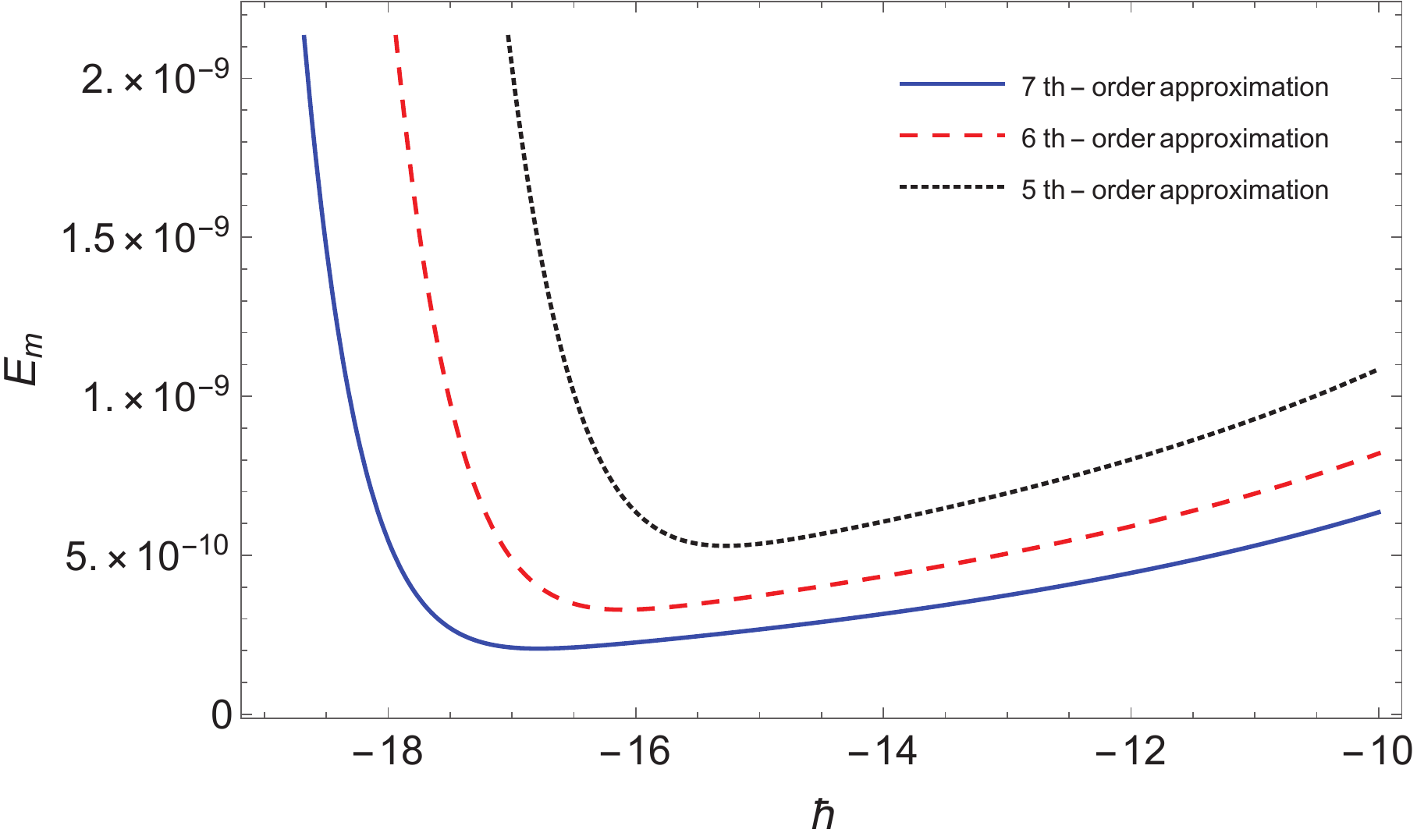}
\caption{Residual error Eq. (\ref{residual}) $E_{m}$ vs the auxiliary parameter $\hbar$ for Run-3 of Coleman \cite{coleman1986effects} experimental data.}\label{Fig:hcurve}
\end{figure}

\begin{figure}[!htb]
\centering
\includegraphics[scale=.5]{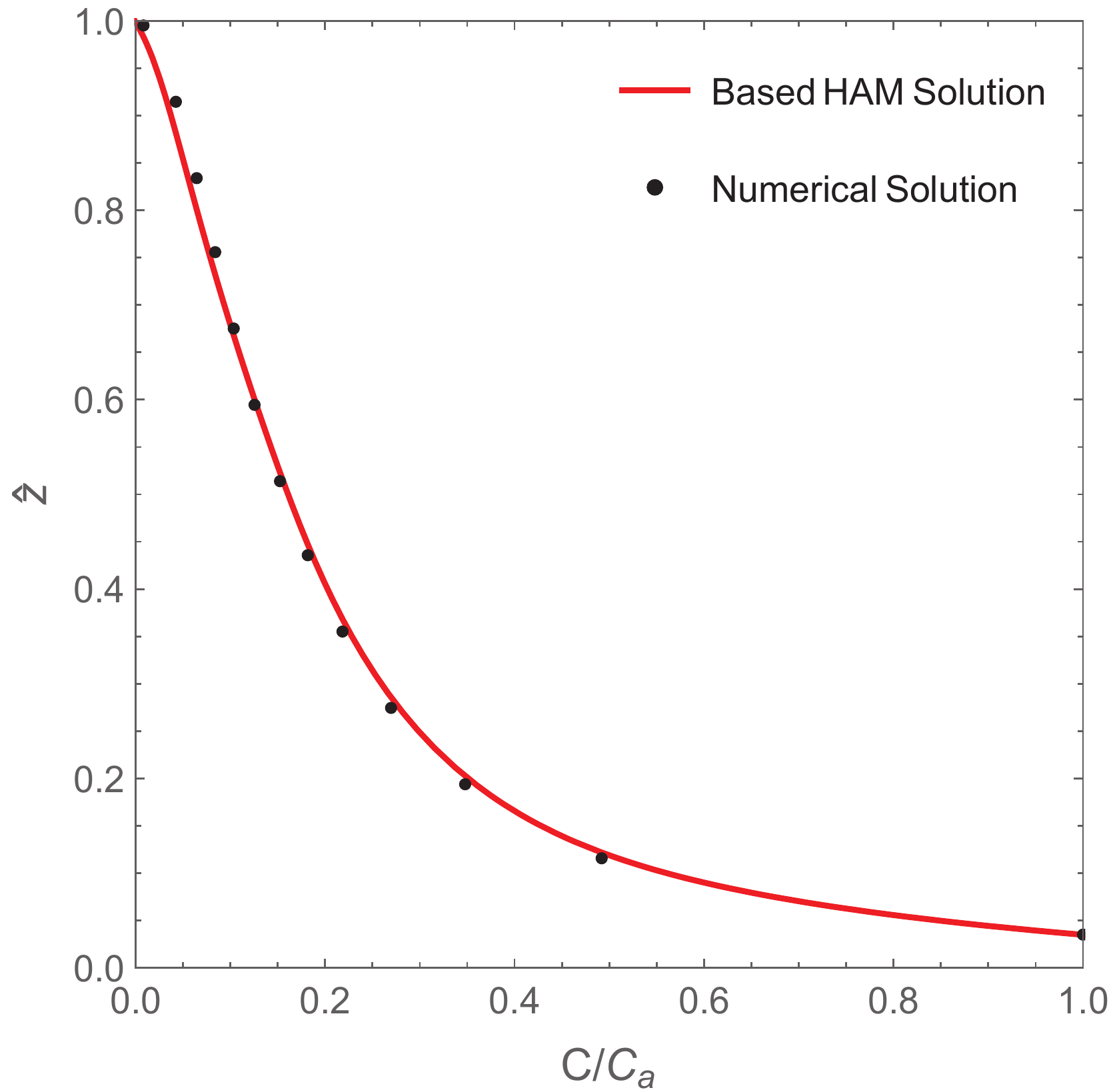}
\caption{Validation of the HAM-based solution (for $\hbar=-16.787$) with numerical solution.}\label{fig :Ham_numerical}
\end{figure}

HAM based series solution Eq. (\ref{ham15}) of Eq. (\ref{hunt3})  is a function of $\hat{z}$ and $\hbar$. As reported in \cite{liao2003beyond}, convergence of HAM based series solution is dependent on $\hbar$ which is the convergence control parameter. Thus to get a convergent solution of our non-linear differential equation, we need a suitable value of $\hbar$. Liao \cite{liao2003beyond} claims that optimal value of convergence control parameter $\hbar$ ensures the convergence of the HAM series in the fastest manner. Liao \cite{liao2003beyond} proposed the classical $\hbar$-level curves method for determination of optimal value of $\hbar$.
Further, Liao \cite{liao2010optimal} introduced a new method for obtaining the optimal value of $\hbar$, called square residual error ($Res(\hbar)$) method given as follows:
\bea
Res(\hbar)=\int_{\hat{a}}^{1}\Bigg\{\mathscr{N}\Big[\sum_{k=0}^{M}C_{k}(\hat{z})\Big]\Bigg\}^{2}d\hat{z}
\eea
Following the work of \cite{liao2010optimal}, we choose the discrete form of above equation as it reduces the computational time. Discrete form of square residual error is given as follows:
\bea
E_m=\frac{1}{L+1}\Bigg(\mathlarger{\sum}_{i=0}^{L}\Bigg\{\mathscr{N}\Big[\sum_{k=0}^{M}C_{k}(\hat{z_{i}})\Big]\Bigg\}^{2}\Bigg) \label{residual}
\eea
where $L+1$ is equally distributed discrete points. We can find the optimal value of $\hbar$ for each order of approximations for which $E_m$ is minimum. In Fig. \ref{Fig:hcurve}, the residual error $E_m$  against the convergence control parameter of $5th$, $6th$ and $7th$ order approximation for Run-3 of Coleman \cite{coleman1986effects} experimental data  is plotted. It can be observed from the figure that error decreases with increase of the order of approximation.
Now we verify the HAM based series solution with numerical solution and for that purpose ``NDSolve" of \emph{Mathematica} is used. We choose the required parameters from Run-3 of Coleman \cite{coleman1986effects} experimental data given in Table \ref{tab:coleman} and the convergence control parameter $\hbar$ is calculated by minimizing the residual error $E_m$ as stated in the previous paragraph. The obtained value for $\hbar$ of $7th$ order approximation is -16.787. An approximate analytical solution of $7th$-order approximation and numerical solution are displayed in Fig \ref{fig :Ham_numerical}. From the figure it can be observed that the HAM based series solution and numerical solution are close to each other which shows the validity of analytical method HAM.

To show the convergence of HAM based series solution, a table that reports numerical and analytical solutions of different orders of approximations is provided. Comparison of numerical solution with $3^{rd}, 4^{th}, 5^{th}, 6^{th}$ and $7^{th}$ order of approximations for $C(\hat{z})$ is shown in Table \ref{tab:hamnumerical}. It may be noted that HAM is an efficient method and not time consuming despite the fact that it is based on the computation involving functions instead of numbers. The time taken by the computer for producing $3^{rd}, 4^{th}, 5^{th}, 6^{th}$ and $7^{th}$ order of approximations of HAM Based solution are 0.23318s, 0.404859s, 0.55446s, 0.866745s and 1.0133s respectively (s-second), which proves its usefulness for practical purposes. All the calculations were performed on a PC having configurations as Intel (R) Core (TM) i5-6500 CPU @ 3.20GHz 64-bit with 4.00 GB of RAM, using Mathematica 11.0. From the table, it can be observed that the solution obtained by HAM is convergent to numerical solution with the increase of each order of approximations.
\FloatBarrier
\begin{table}[H]
  \centering
  \caption{Comparison of the numerical solution with the approximations of $C(\hat{z})$}
    \begin{tabular}{clrrrrr}
    \hline
    \multicolumn{1}{c}{$\hat{z}$} & \begin{tabular}[c]{@{}l@{}l@{}l@{}}Numerical\\ Solution\end{tabular} & \begin{tabular}[c]{@{}l@{}}7th order\\ approx.\end{tabular}  & \begin{tabular}[c]{@{}l@{}}6th order\\ approx.\end{tabular} & \begin{tabular}[c]{@{}l@{}}5th order\\ approx.\end{tabular} & \begin{tabular}[c]{@{}l@{}}4th order\\ approx.\end{tabular} & \begin{tabular}[c]{@{}l@{}}3th order\\ approx.\end{tabular} \\ \hline
    0.035 & 0.0017000 & 0.0017000 & 0.0017000 & 0.0017000 & 0.0017000 & 0.0017000 \\
    0.085 & 0.0010073 & 0.0010588 & 0.0010794 & 0.0011316 & 0.0011971 & 0.0012927 \\
    0.135 & 0.0007541 & 0.0007811 & 0.0007925 & 0.0008286 & 0.0008863 & 0.0009944 \\
    0.185 & 0.0006117 & 0.0006303 & 0.0006367 & 0.0006570 & 0.0006937 & 0.0007796 \\
    0.235 & 0.0005165 & 0.0005316 & 0.0005361 & 0.0005490 & 0.0005705 & 0.0006271 \\
    0.285 & 0.0004465 & 0.0004593 & 0.0004629 & 0.0004727 & 0.0004867 & 0.0005196 \\
    0.335 & 0.0003917 & 0.0004029 & 0.0004059 & 0.0004138 & 0.0004246 & 0.0004432 \\
    0.385 & 0.0003468 & 0.0003568 & 0.0003594 & 0.0003659 & 0.0003749 & 0.0003874 \\
    0.435 & 0.0003089 & 0.0003177 & 0.0003206 & 0.0003255 & 0.0003332 & 0.0003442 \\
    0.485 & 0.0002760 & 0.0002833 & 0.0002883 & 0.0002899 & 0.0002974 & 0.0003081 \\
    0.535 & 0.0002467 & 0.0002518 & 0.0002622 & 0.0002575 & 0.0002668 & 0.0002753 \\
    0.585 & 0.0002202 & 0.0002222 & 0.0002416 & 0.0002268 & 0.0002407 & 0.0002439 \\
    0.635 & 0.0001957 & 0.0001939 & 0.0002256 & 0.0001973 & 0.0002181 & 0.0002128 \\
    0.685 & 0.0001727 & 0.0001669 & 0.0002124 & 0.0001689 & 0.0001981 & 0.0001819 \\
    0.735 & 0.0001507 & 0.0001413 & 0.0002002 & 0.0001415 & 0.0001795 & 0.0001515 \\
    0.785 & 0.0001292 & 0.0001171 & 0.0001875 & 0.0001150 & 0.0001610 & 0.0001216 \\
    0.835 & 0.0001076 & 0.0000938 & 0.0001730 & 0.0000893 & 0.0001417 & 0.0000922 \\
    0.885 & 0.0000850 & 0.0000706 & 0.0001555 & 0.0000633 & 0.0001204 & 0.0000621 \\
    0.935 & 0.0000599 & 0.0000456 & 0.0001330 & 0.0000350 & 0.0000948 & 0.0000291 \\
    0.985 & 0.0000256 & 0.0000125 & 0.0001002 & -0.0000010 & 0.0000602 & -0.0000106 \\
    \hline
    \begin{tabular}[c]{@{}c@{}}Running time\\ in second\end{tabular} & \multicolumn{1}{c}{-} & 1.0133000& 0.8667450 & 0.5544600 & 0.4000000 & 0.2331800 \\
   \hline
    \end{tabular}%
 \label{tab:hamnumerical}%
\end{table}%
\FloatBarrier
\subsection{Physical interpretation}
\subsubsection{Comparison of shear-induced and sediment diffusion coefficient}
\begin{figure}[!htb]
\centering
\includegraphics[scale=.6]{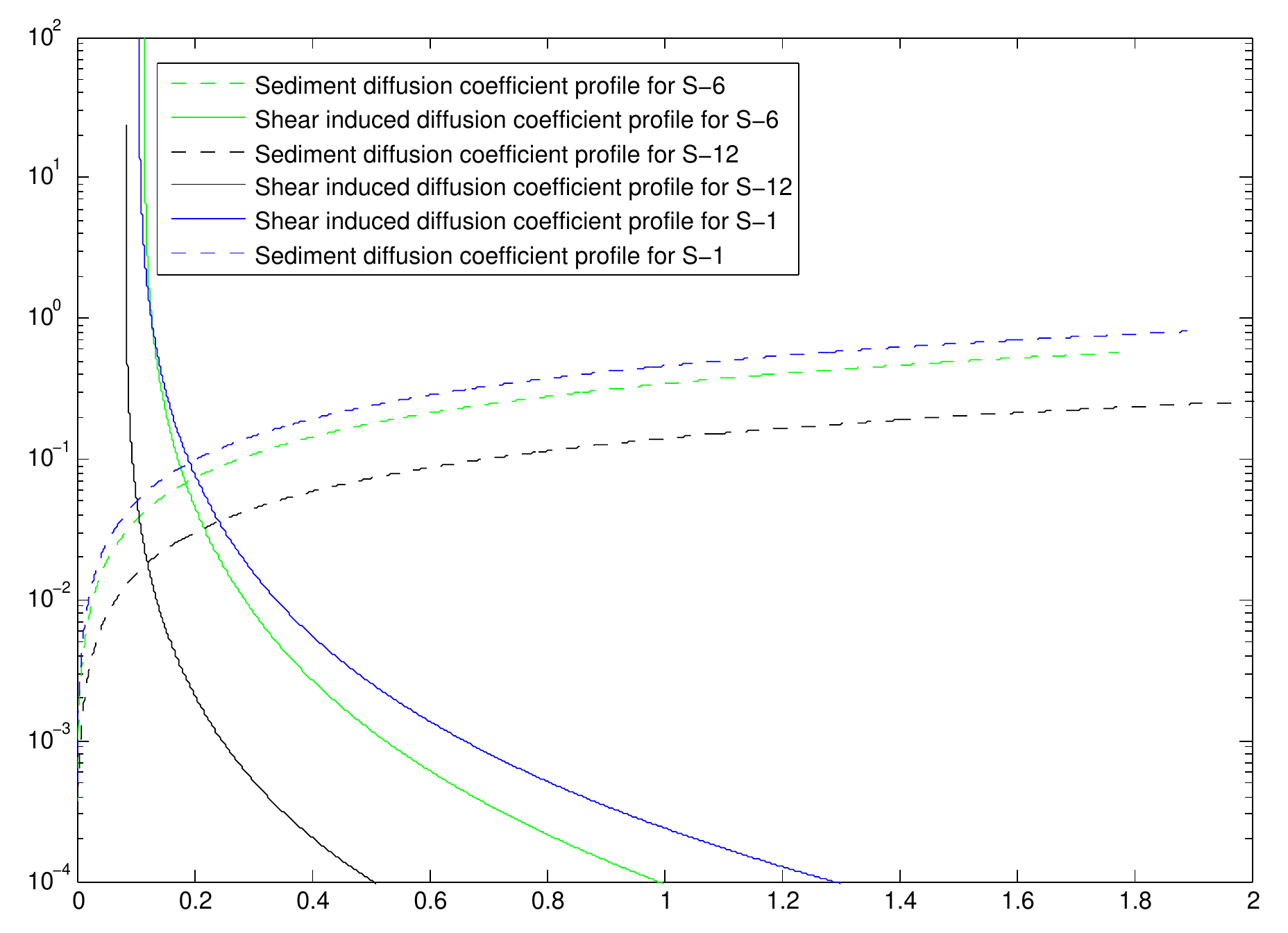}
\caption{Comparison of shear-induced and sediment diffusion coefficient.}
\label{fig:Shear}
\end{figure}

Profiles of shear-induced diffusion coefficient ($\varepsilon_{cg}+\varepsilon_{sg}$) and sediment diffusion against the vertical distance $z$ are plotted in Fig. \ref{fig:Shear}. Three experimental runs ($S1,~S6$ and $S12$) from the data of Einstein and Chien \cite{einstein1955effects} are used to compare the shear-induced and sediment diffusion coefficient. It can be observed from Eqs. (\ref{CG1}) and (\ref{SG}) that to assess the profiles of $\varepsilon_{cg}$ or $\varepsilon_{sg}$, one needs a concentration profile. In our study the widely known Rouse equation has been used for this purpose. It can be observed from the figure that the magnitude of shear-induced diffusion coefficient is relatively higher near the bed than the rest of the flow depth and takes on vanishingly small values at the upper portion of the flow depth. For a fixed distance $z$, effect of shear-induced diffusion coefficient is more for large particles as compared to the smaller particles. On the other hand, effect of sediment diffusion coefficient is negligible near the bed and increases with the increase of distance from the bed. Moreover, for a given $z$, a small variation of sediment diffusion coefficient is observed for varying particle diameter. From Fig. \ref{fig:Shear}, it is concluded that effect of shear-induced diffusion coefficient is prominent near the bed as compared to the sediment diffusion coefficient, and changes with the size of particle diameter.
\subsubsection{Shear-induced diffusion coefficient against shear gradient and shear-induced diffusion coefficient against concentration gradient}
\begin{figure}[!htb]
  \centering
  \subfloat[Profile of $\varepsilon_{sg}$ for different diameters of bed particles.]{\includegraphics[scale=0.5]{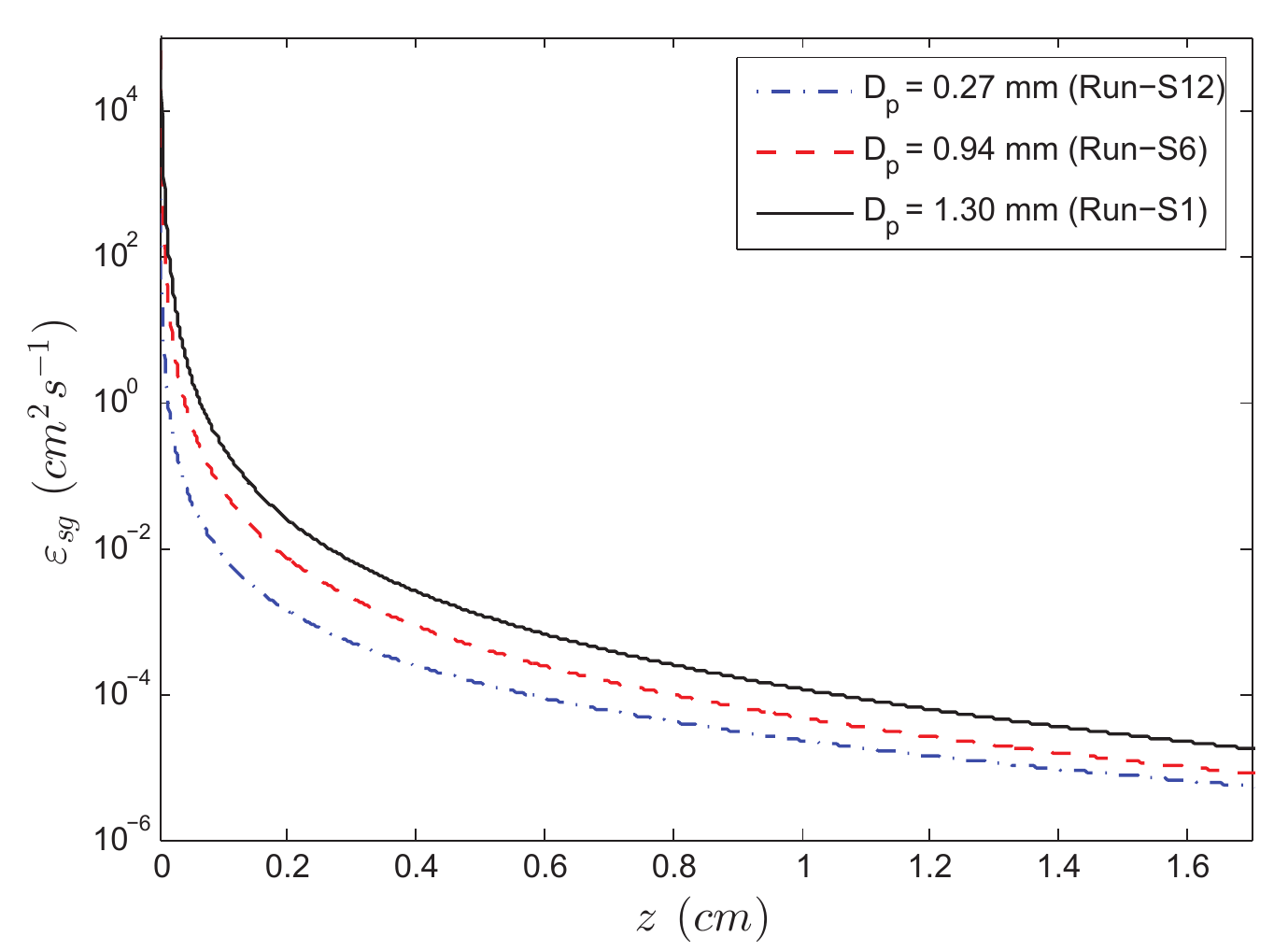}}
  \hfill
  \subfloat[Profile of $\varepsilon_{cg}$ for different diameters of bed particles.]{\includegraphics[scale=0.5]{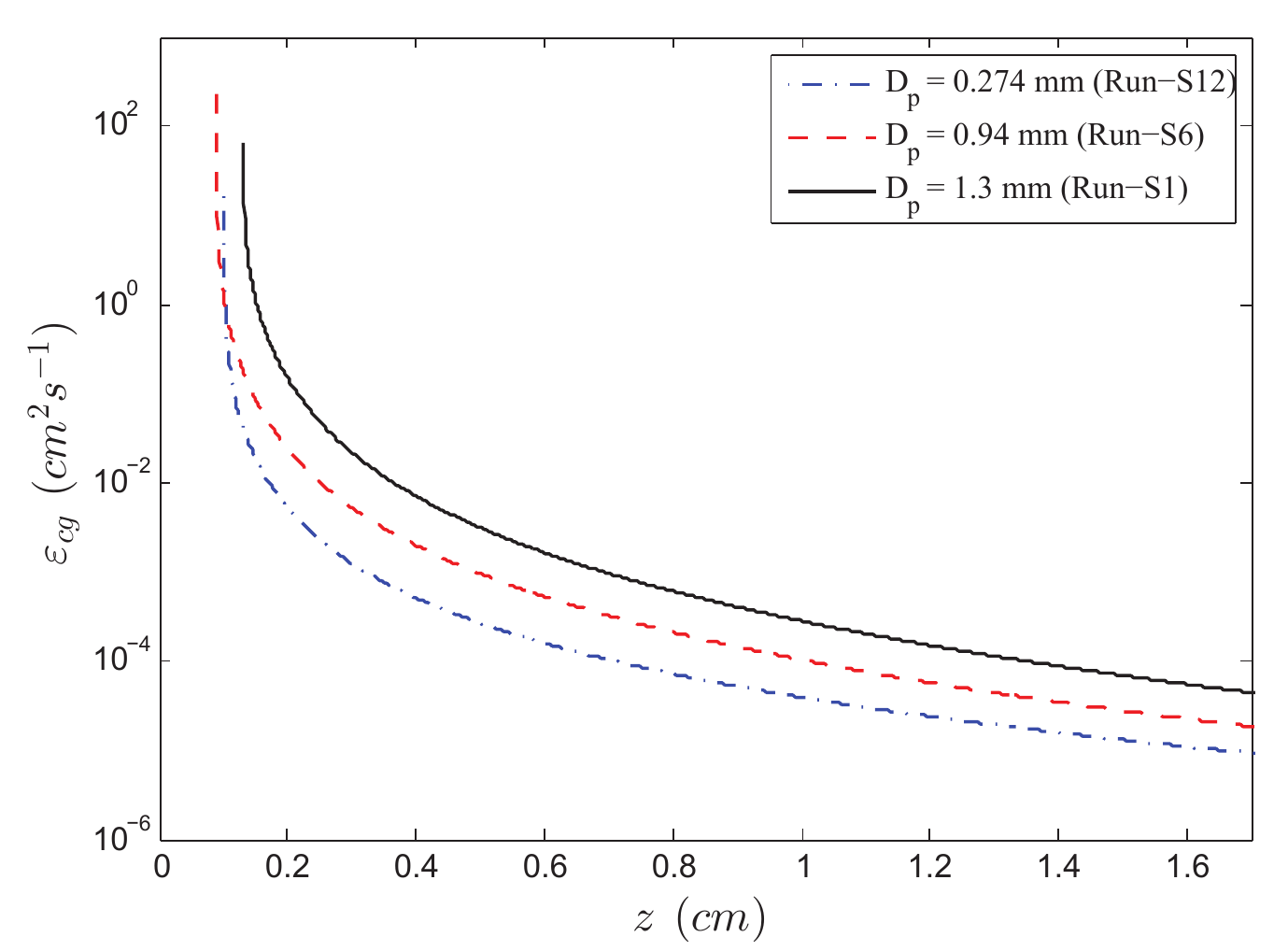}}
  \hfill
  \subfloat[Comparison of $\varepsilon_{sg}$ and $\varepsilon_{cg}$ for different diameter of bed particles.]{\includegraphics[scale=0.5]{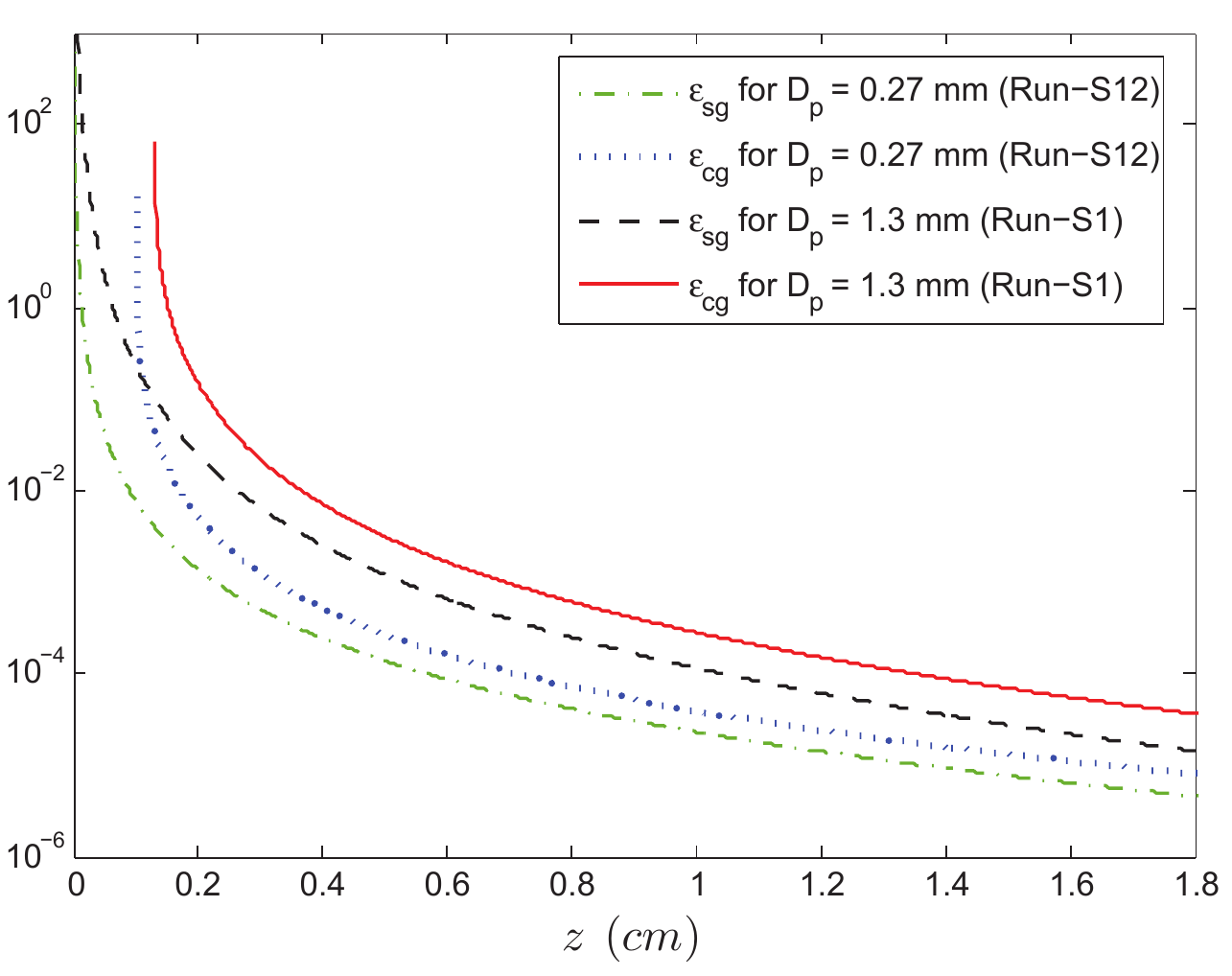}}
  \hfill
  \caption{Profile of $\varepsilon_{sg}$ and $\varepsilon_{cg}$ for different diameter of bed particles along with distance from the bed.} \label{fig:shear_sediment}
  \end{figure}

Fig. \ref{fig:shear_sediment} (a) shows the variation of shear-induced diffusion coefficient against shear gradient with vertical distance $z$ and varying particle diameter for three experimental runs ($S1,~S6~\mbox{and}~S12$) from the data of Einstein and Chien \cite{einstein1955effects}. It can be observed from the figure that magnitude of $\varepsilon_{sg}$ decreases with the increase of vertical distance, and for a fixed distance say $z=z_0$, it is more for larger particles.
\par
Fig. \ref{fig:shear_sediment} (b) shows the profiles of shear-induced diffusion coefficient against concentration gradient $\varepsilon_{cg}$ for three experimental runs ($S1,~S6~\mbox{and}~S12$) from the data of Einstein and Chien \cite{einstein1955effects}. From the figure, it can be noticed that $\varepsilon_{cg}$ follows same characteristics as that of  $\varepsilon_{sg}$.
\par
Profiles of  shear-induced diffusion coefficient against concentration gradient $\varepsilon_{sg}$ and shear-induced diffusion coefficient against concentration gradient $\varepsilon_{cg}$ are plotted in Fig. \ref{fig:shear_sediment} (c) for two experimental runs ($S1 \mbox{ and }S12$) from the data of Einstein and Chien \cite{einstein1955effects}. It can be seen from the figure that the effect of  $\varepsilon_{cg}$ is more than effect of  $\varepsilon_{sg}$ for a fixed particle diameter and a particular height.
\subsubsection{Variation of concentration profile with $\varepsilon$, $\beta$ and $\eta$}
\begin{figure}[!htb]
\centering
\includegraphics[scale=.5]{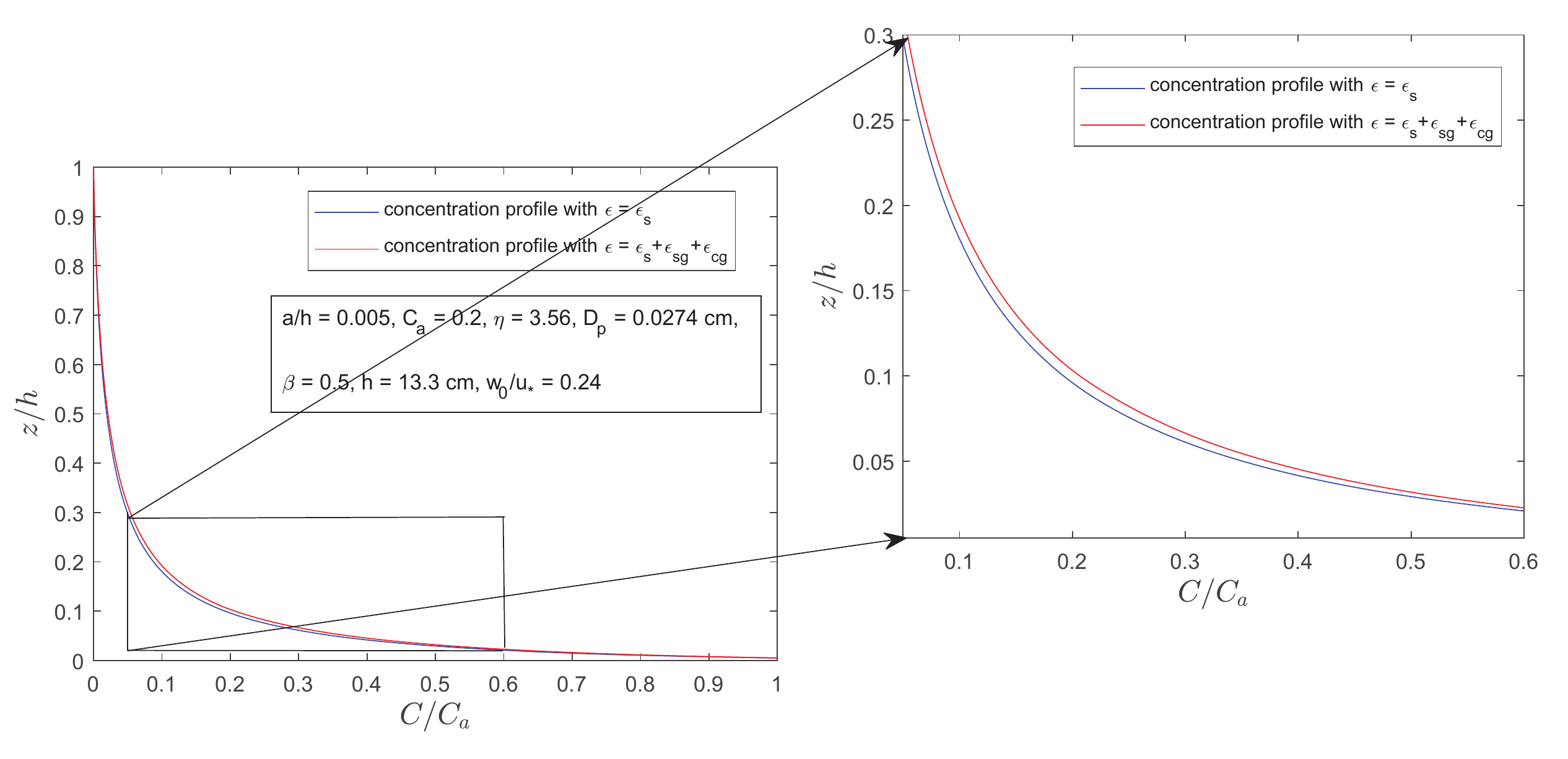}
\caption{Concentration profile of sediment with diffusion coefficient as total diffusion coefficient and diffusion coefficient as sediment diffusion coefficient}\label{with-without}
\end{figure}

\begin{figure}[!htb]
\centering
\includegraphics[scale=.6]{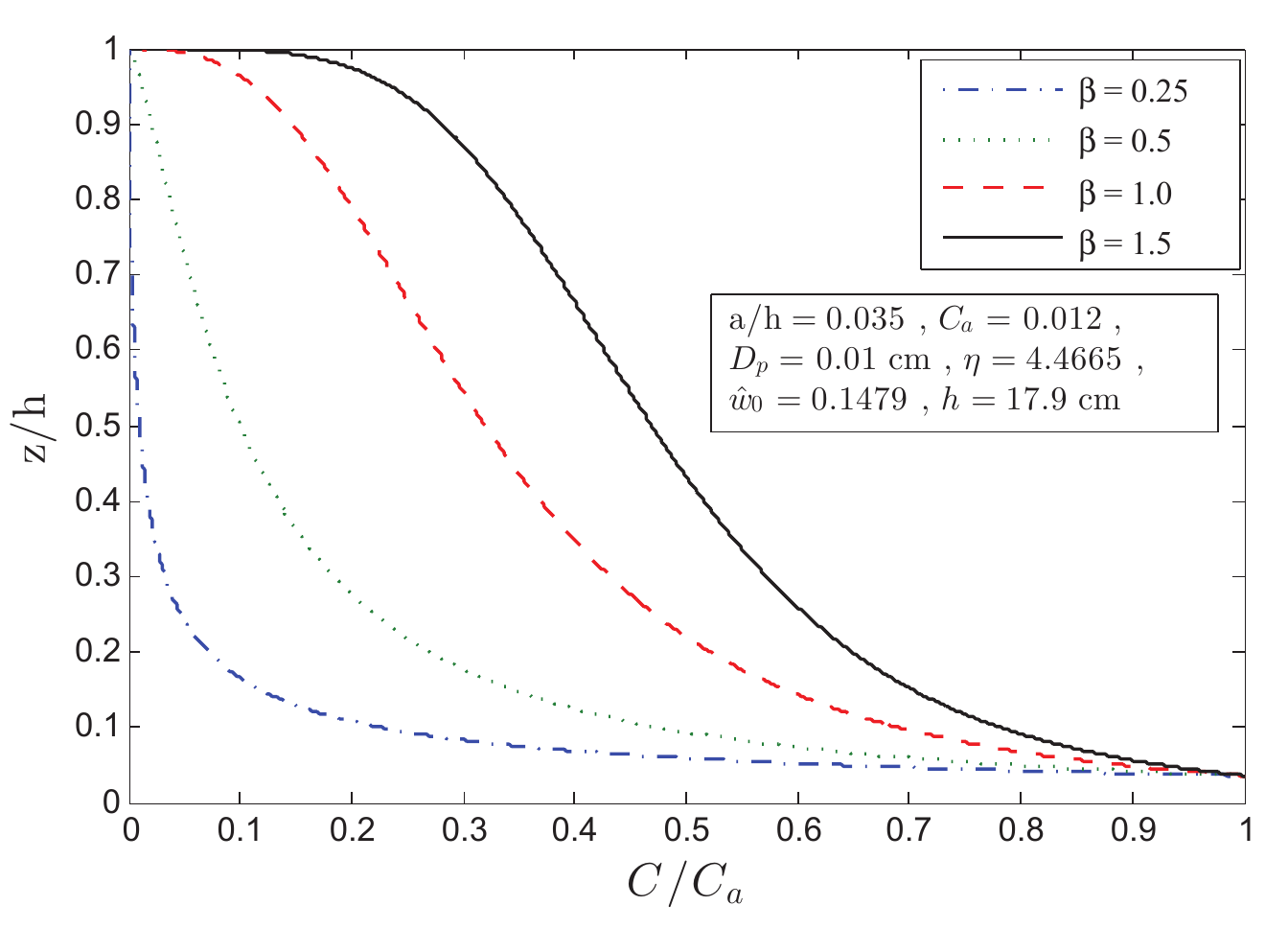}
\caption{Vertical concentration profiles for different values of $\beta$.}\label{fig :beta}
\end{figure}

\begin{figure}[!htb]
\centering
\includegraphics[scale=.6]{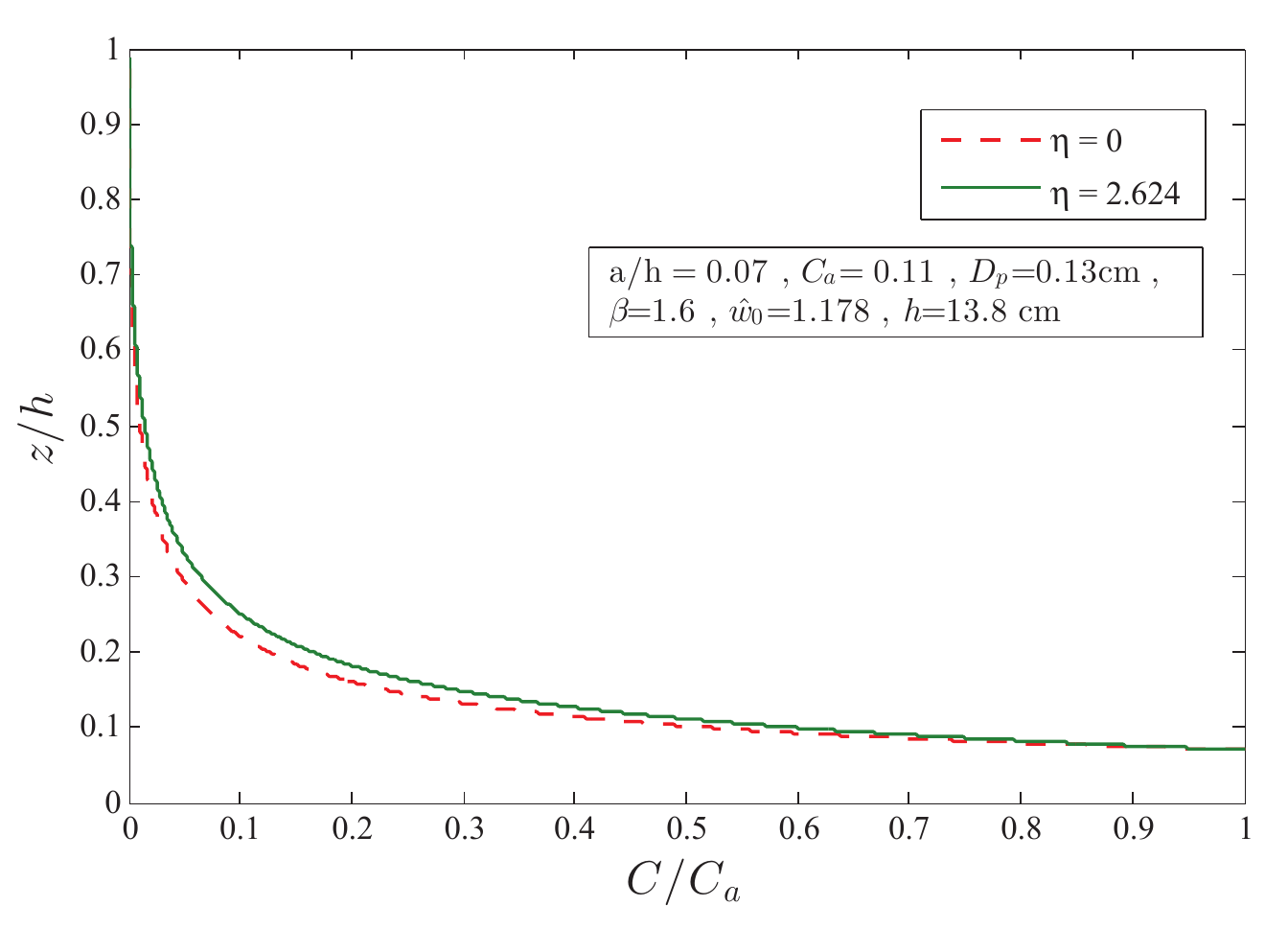}
\caption{Effect of $\eta$ on the vertical concentration of suspended particles.}\label{fig :eta}
\end{figure}

Two concentration profiles of suspended particle with and without the inclusion of shear-induced diffusion coefficient in total diffusion coefficient are displayed in Fig. \ref{with-without}. It is observed from Fig. \ref{with-without} that the concentration values differ, mainly near the channel bed, as already discussed that the effect of shear-induced diffusion coefficient is prominent thereat. For clear understanding, the near-bed region where the effect is visible, is shown in an enlarged way in the same figure.
Vertical concentration profiles of suspended particles for different values of $\beta$ are presented in Fig. \ref{fig :beta}. The required parameters are used from Run-11 of Coleman \cite{coleman1986effects} experimental data (See Table \ref{tab:coleman}). It can be observed from Fig. \ref{fig :beta} that concentration profile is sensitive to the values of $\beta$. A lower value of $\beta$ indicates a concave profile while higher $\beta$ produces convex profile. Moreover, at a particular distance say $z=z_0$ from bed, the concentration values are higher for larger $\beta$.
Effect of hindered settling velocity $(w_{s})$ on the vertical concentration profile is depicted in Fig. \ref{fig :eta}. Two profiles of sediment concentration with and without the effect of hindered settling are displayed in the figure. It can be observed from the figure that the values of concentration increase due to the presence of hindered settling velocity. Effect of hindered settling is visible only in the main flow region but not near the water surface and near the bed since the particles are not mostly present near the water surface and not in suspension near the bed .

\subsection{Comparison between present model with Rouse \cite{rouse1937modern} and Hunt \cite{hunt1954turbulent} model}
\begin{figure}[!htb]
  \centering
  \subfloat[]{\includegraphics[scale=0.51]{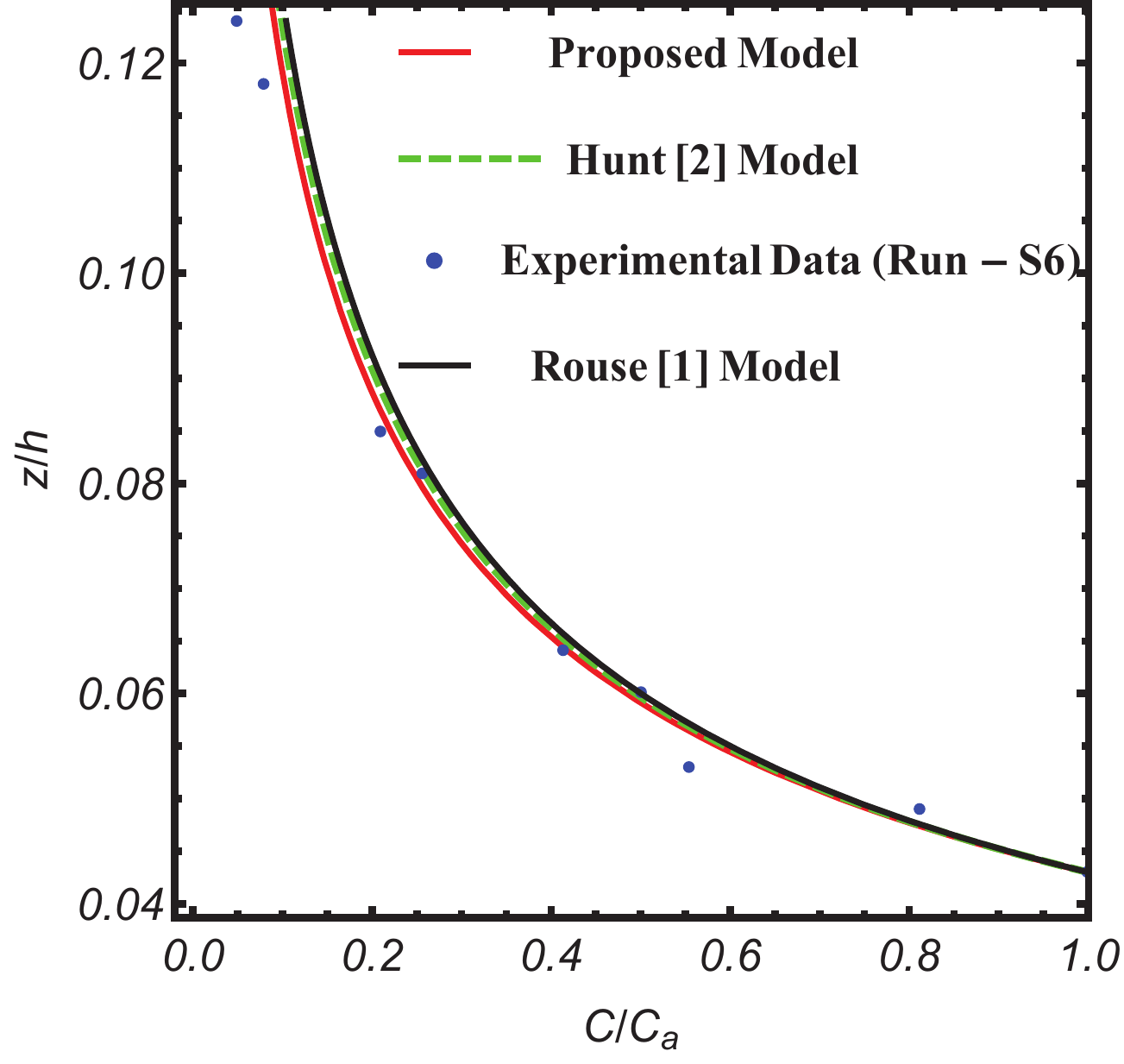}}
  \hfill
  \subfloat[]{\includegraphics[scale=0.5]{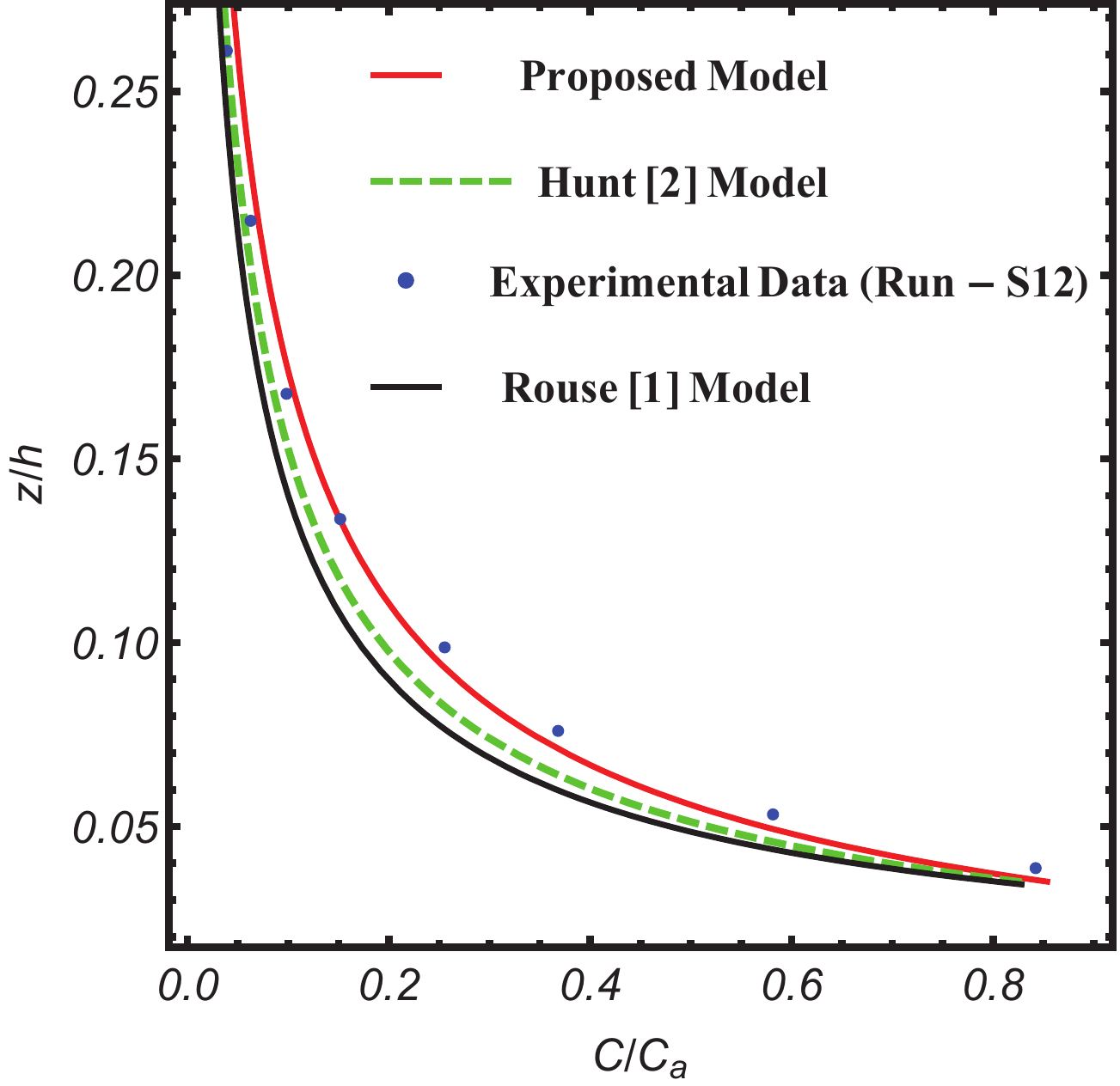}}
  \hfill
  \caption{Comparison between proposed concentration profile, Rouse \cite{rouse1937modern} and Hunt \cite{hunt1954turbulent} model with observed data of Einstein and Chien \cite{einstein1955effects} (a) Run-S6, (b)Run-S12} \label{Rouse_Hunt}
  \end{figure}

In this section, we compare the proposed model with the models of Rouse \cite{rouse1937modern} and Hunt \cite{hunt1954turbulent} and experimental data  to check the accuracy of our model. Three concentration profiles with experimental data of Einstein and Chien \cite{einstein1955effects} for Run-S6 and Run-S12 are plotted in Fig. \ref{Rouse_Hunt}(a) and \ref{Rouse_Hunt}(b) respectively. Difference  between proposed model, Rouse \cite{rouse1937modern} and Hunt  \cite{hunt1954turbulent} model can be easily noticed in Fig. \ref{Rouse_Hunt}, especially in Fig.  \ref{Rouse_Hunt}(b), which is of higher concentration. Due to the presence of hindered  settling effect, shear-induced  diffusion and $\beta \neq 1$, the proposed model is more close to the experimental data than  the models of Rouse \cite{rouse1937modern} and Hunt \cite{hunt1954turbulent} as can be seen from Fig. \ref{Rouse_Hunt}.

\subsection{Comparison between present modified Hunt equation with experimental data}
\begin{figure}[!htb]
  \centering
  \subfloat[]{\includegraphics[scale=0.33]{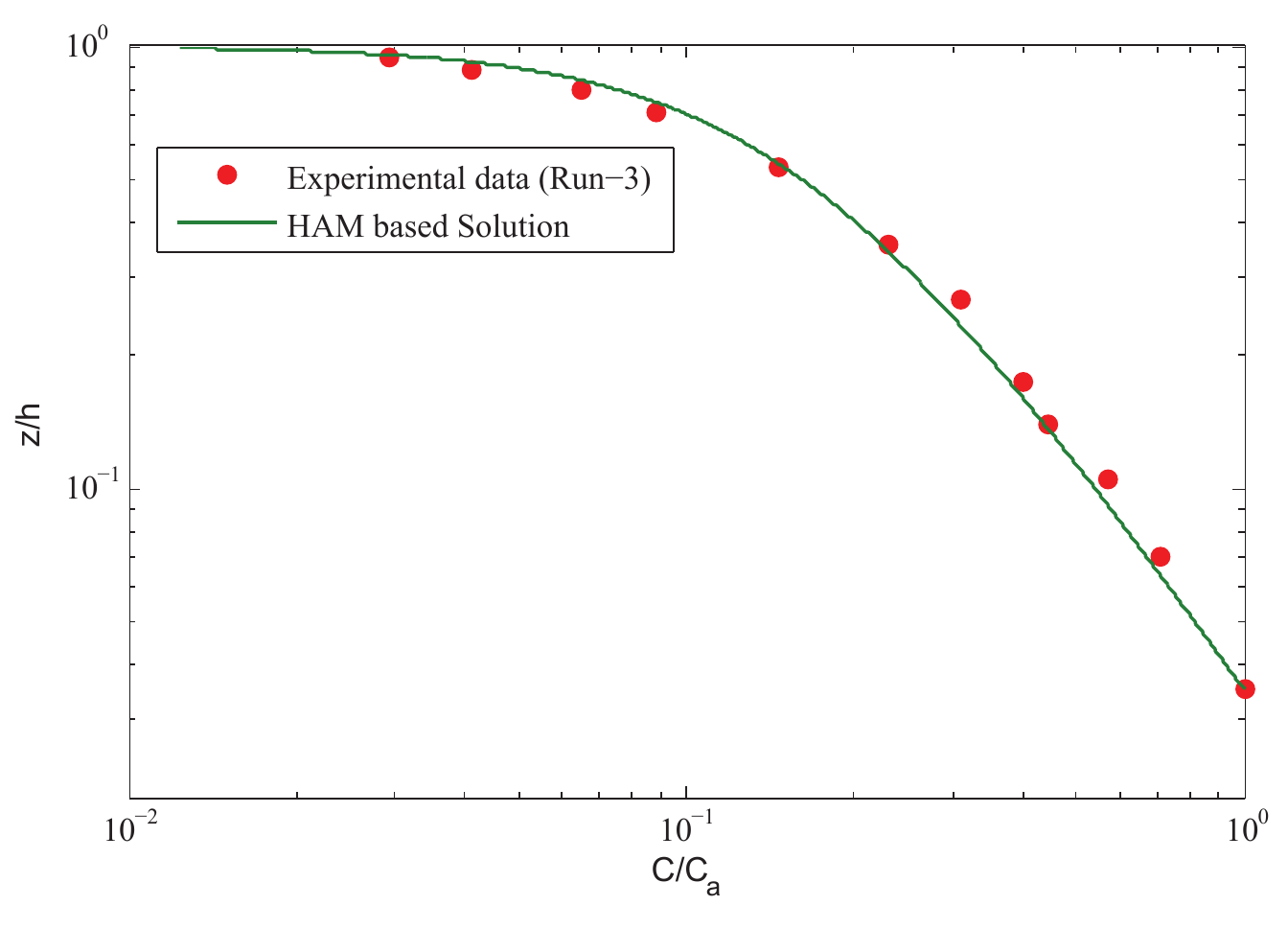}}
  \hfill
  \subfloat[]{\includegraphics[scale=0.33]{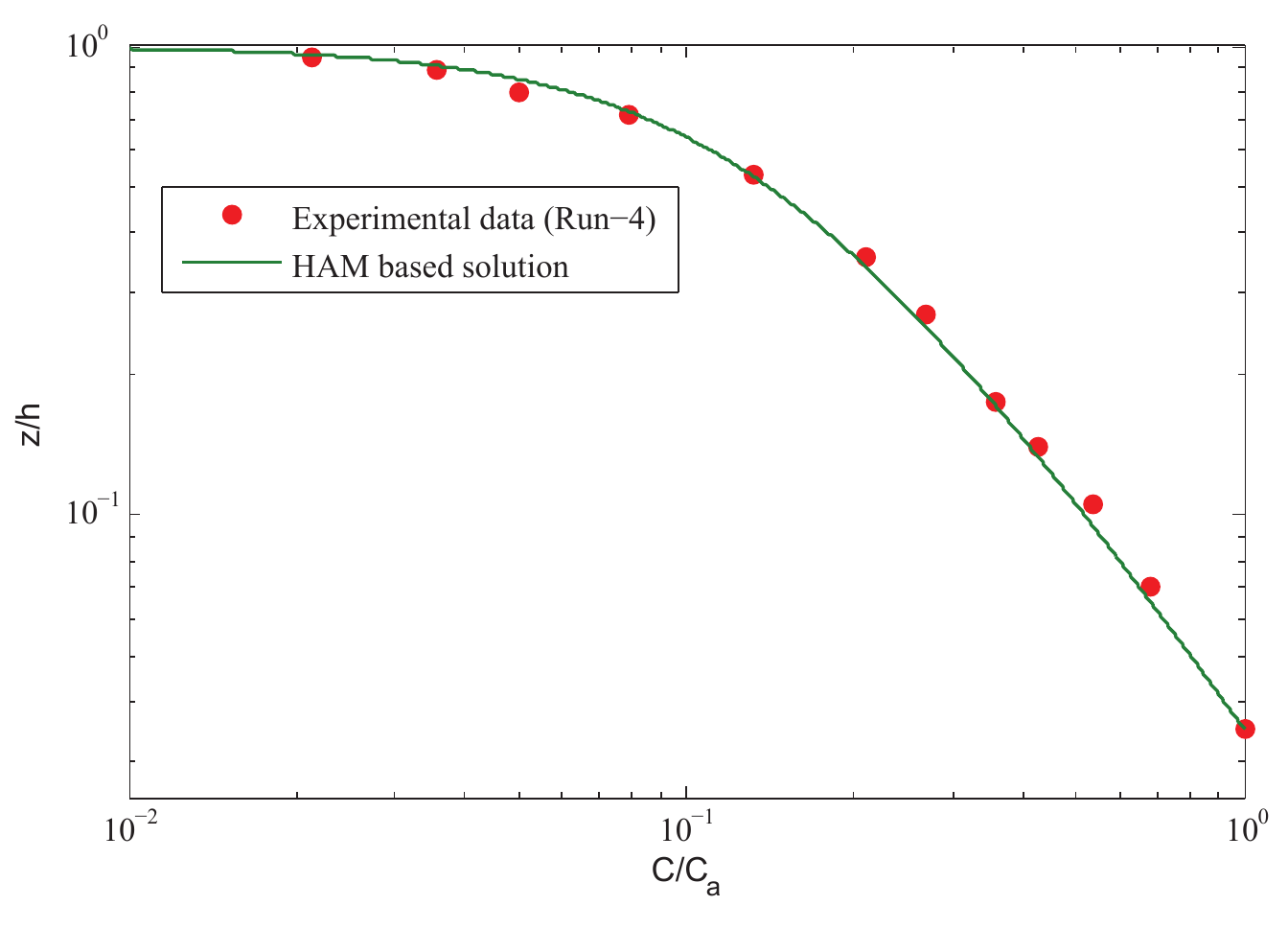}}
  \hfill
  \subfloat[]{\includegraphics[scale=0.33]{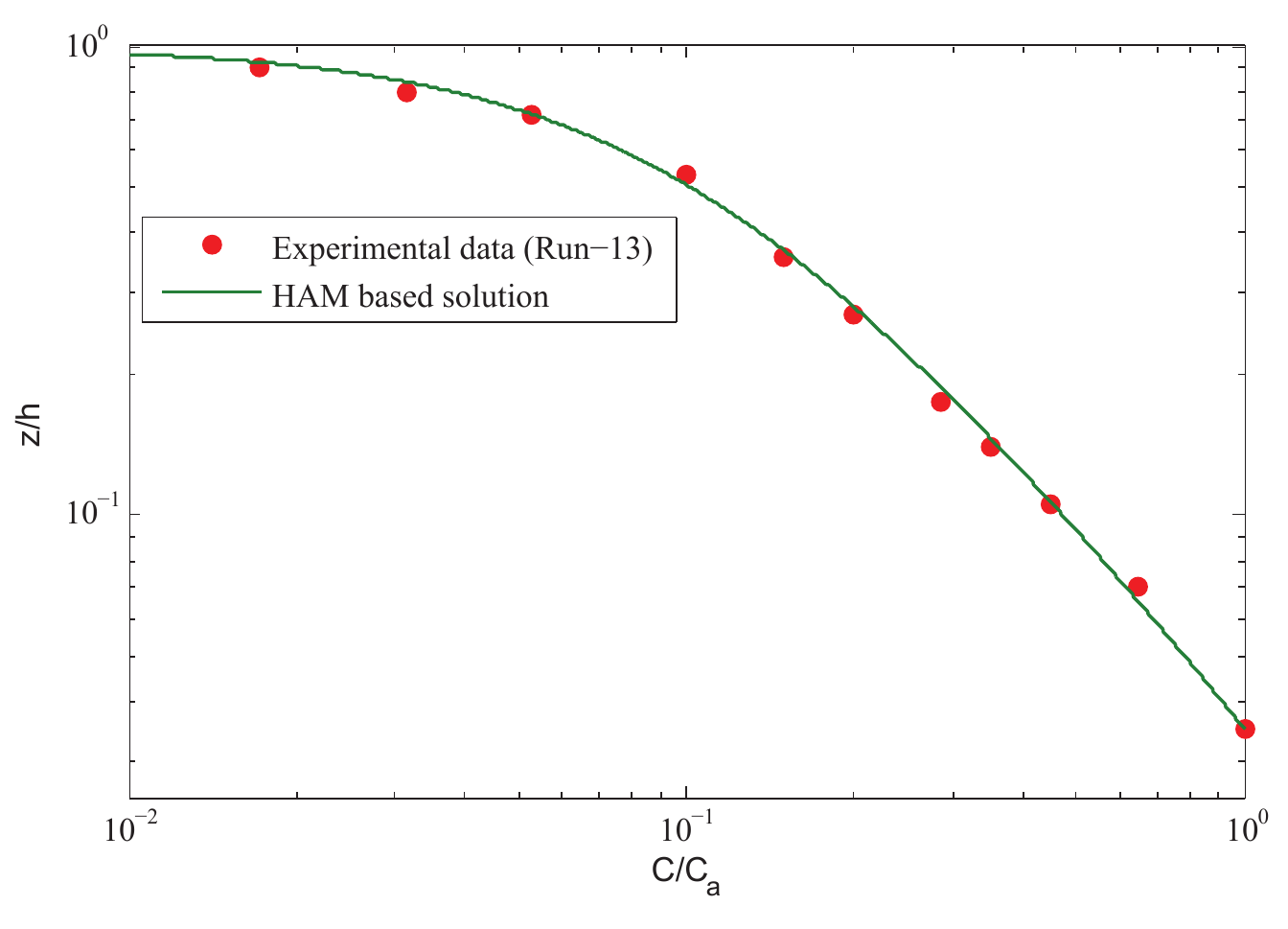}}
  \hfill
  \caption{Comparison between computed particle concentration profile and observed data of Coleman \cite{coleman1986effects} (a) Run-3 , (b) Run-4 and  (c) Run-13} \label{fig:Col}
  \end{figure}

\begin{figure}[!htb]
  \centering
  \subfloat[]{\includegraphics[scale=0.33]{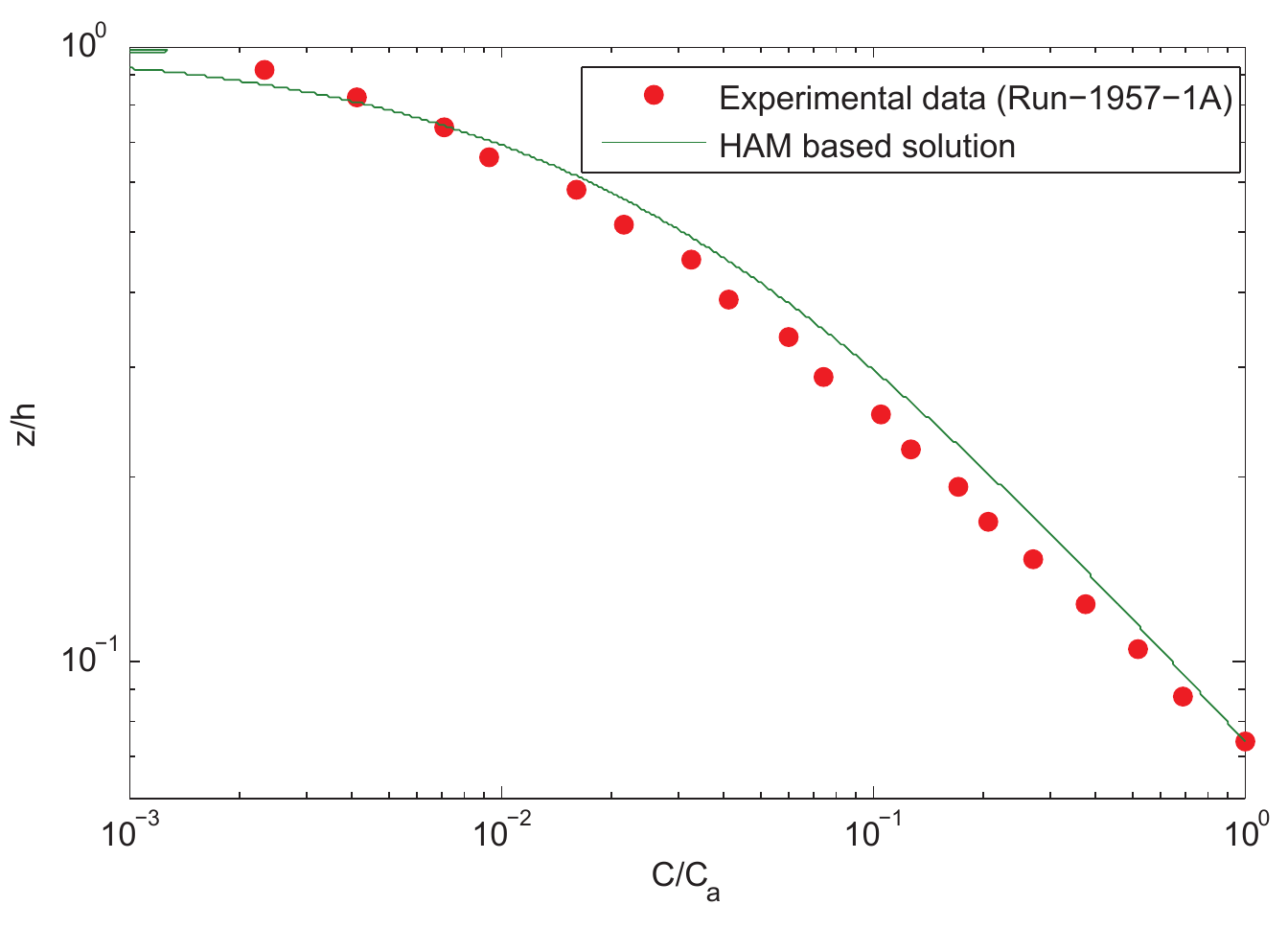}}
  \hfill
  \subfloat[]{\includegraphics[scale=0.33]{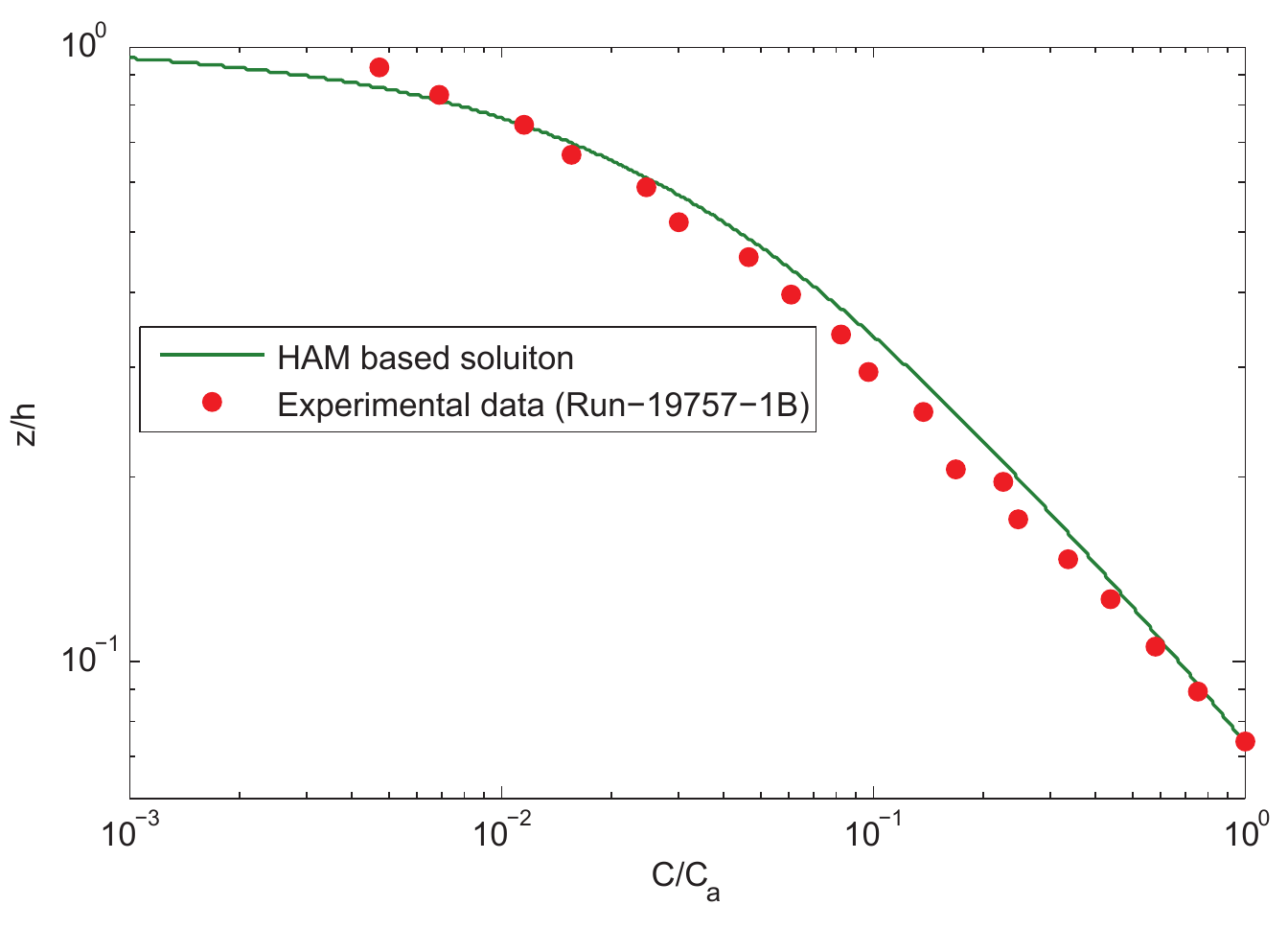}}
  \hfill
  \subfloat[]{\includegraphics[scale=0.33]{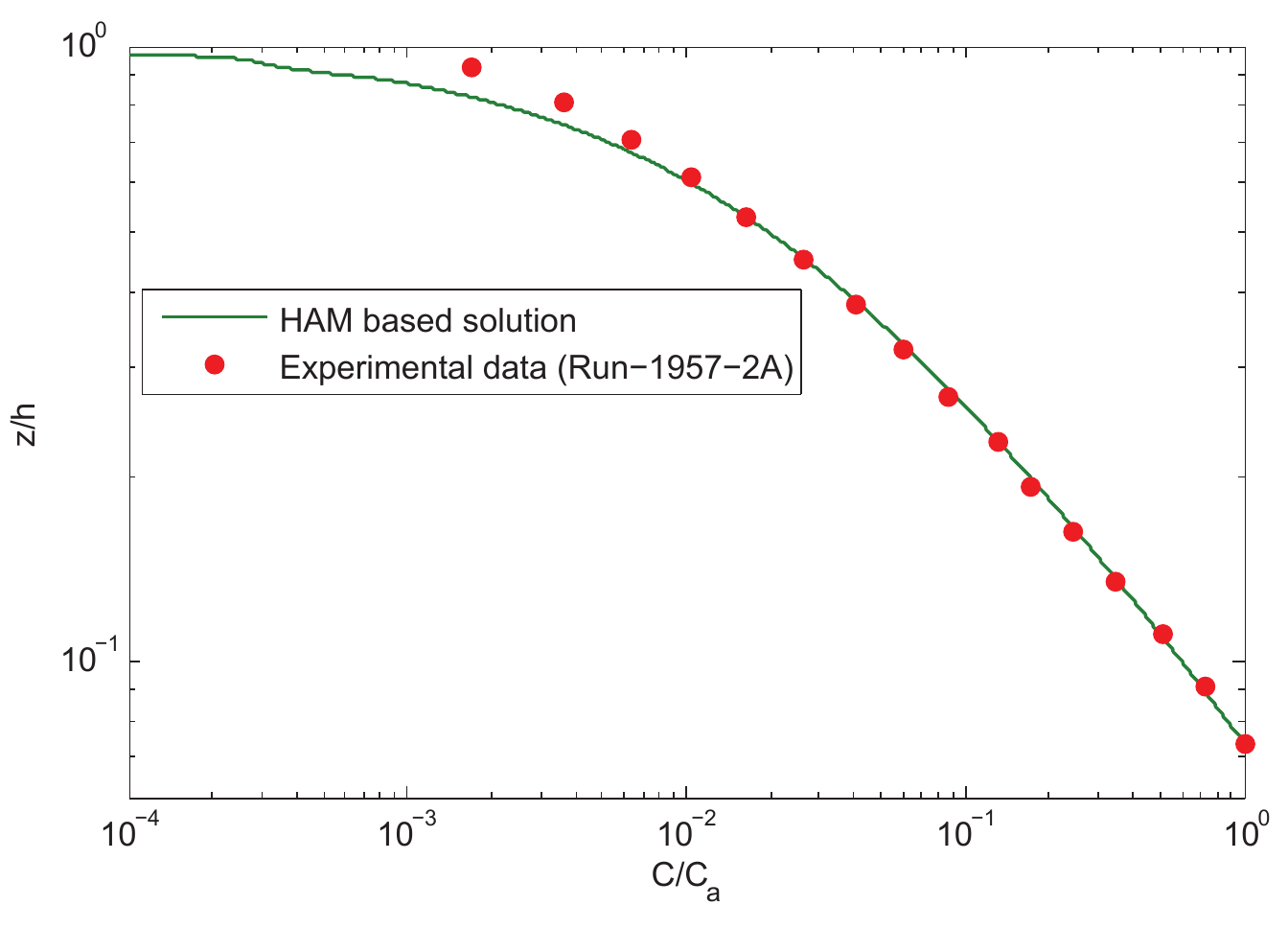}}
  \hfill
  \caption{Comparison between computed particle concentration profile and observed data of Lyn \cite{lyn1988similarity} (a) Run-$1957$ST-$1$A , (b) Run-$1957$ST-$1$B and (c) Run-$1957$ST-$2$A} \label{fig:lyn}
  \end{figure}

\begin{figure}[!htb]
  \centering
  \subfloat[]{\includegraphics[scale=0.33]{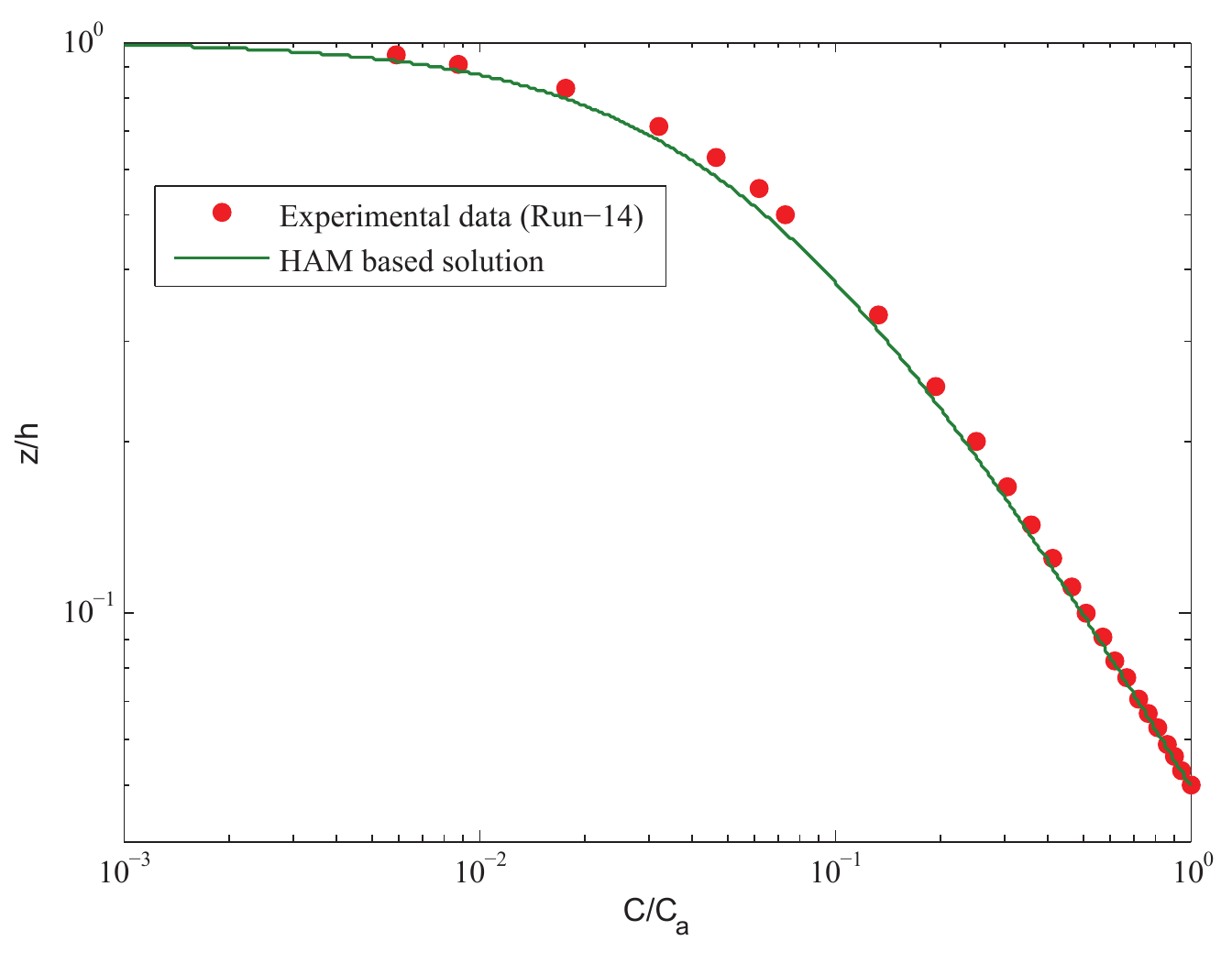}}
  \hfill
  \subfloat[]{\includegraphics[scale=0.33]{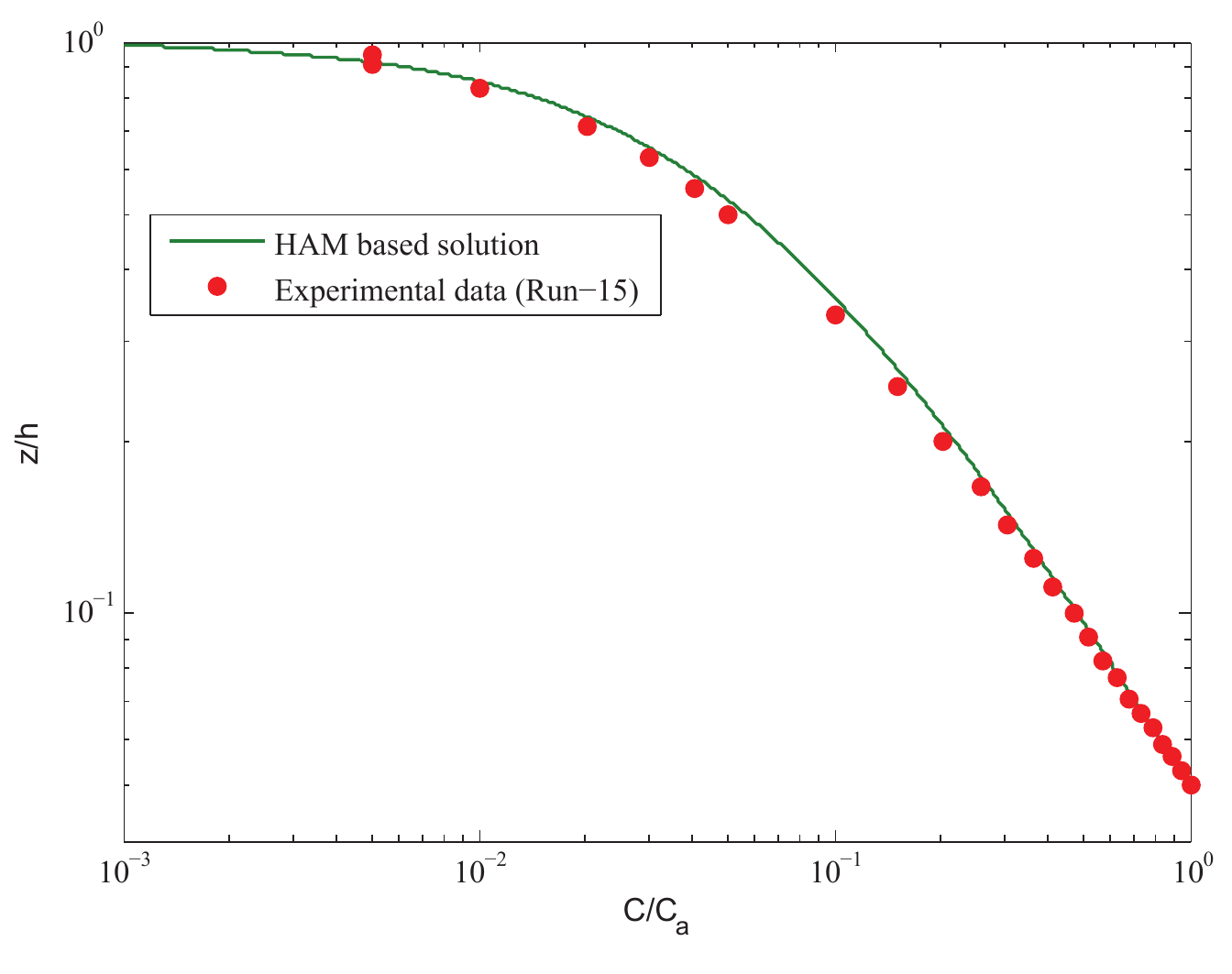}}
  \hfill
  \subfloat[]{\includegraphics[scale=0.33]{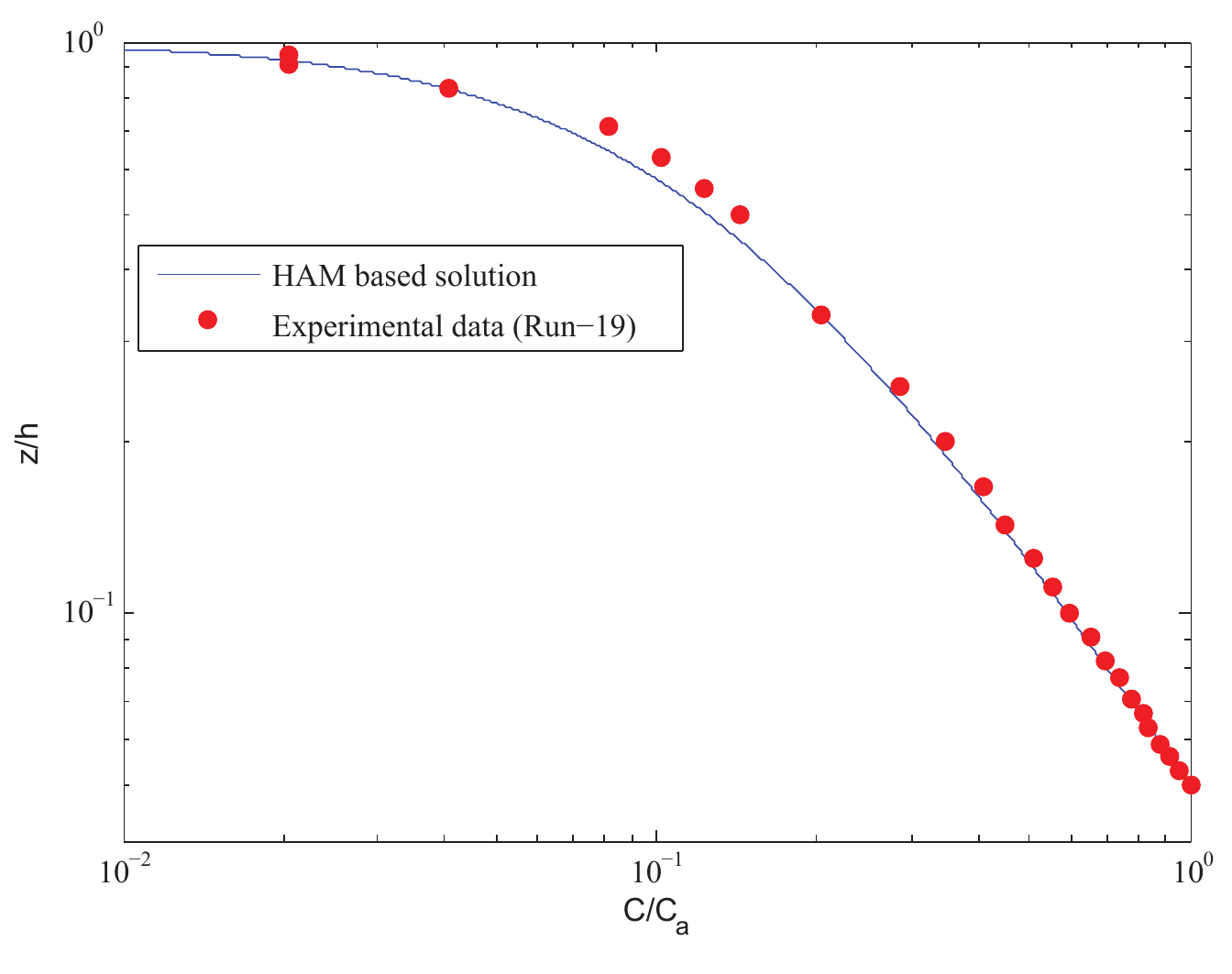}}
  \hfill
  \caption{Comparison between computed particle concentration profile and observed data of Vanoni \cite{vanoni1946transportation} (a) Run-14, (b) Run-15 and (c) Run-19} \label{fig:Vanoni}
  \end{figure}

\begin{figure}[!htb]
  \centering
  \subfloat[]{\includegraphics[scale=0.33]{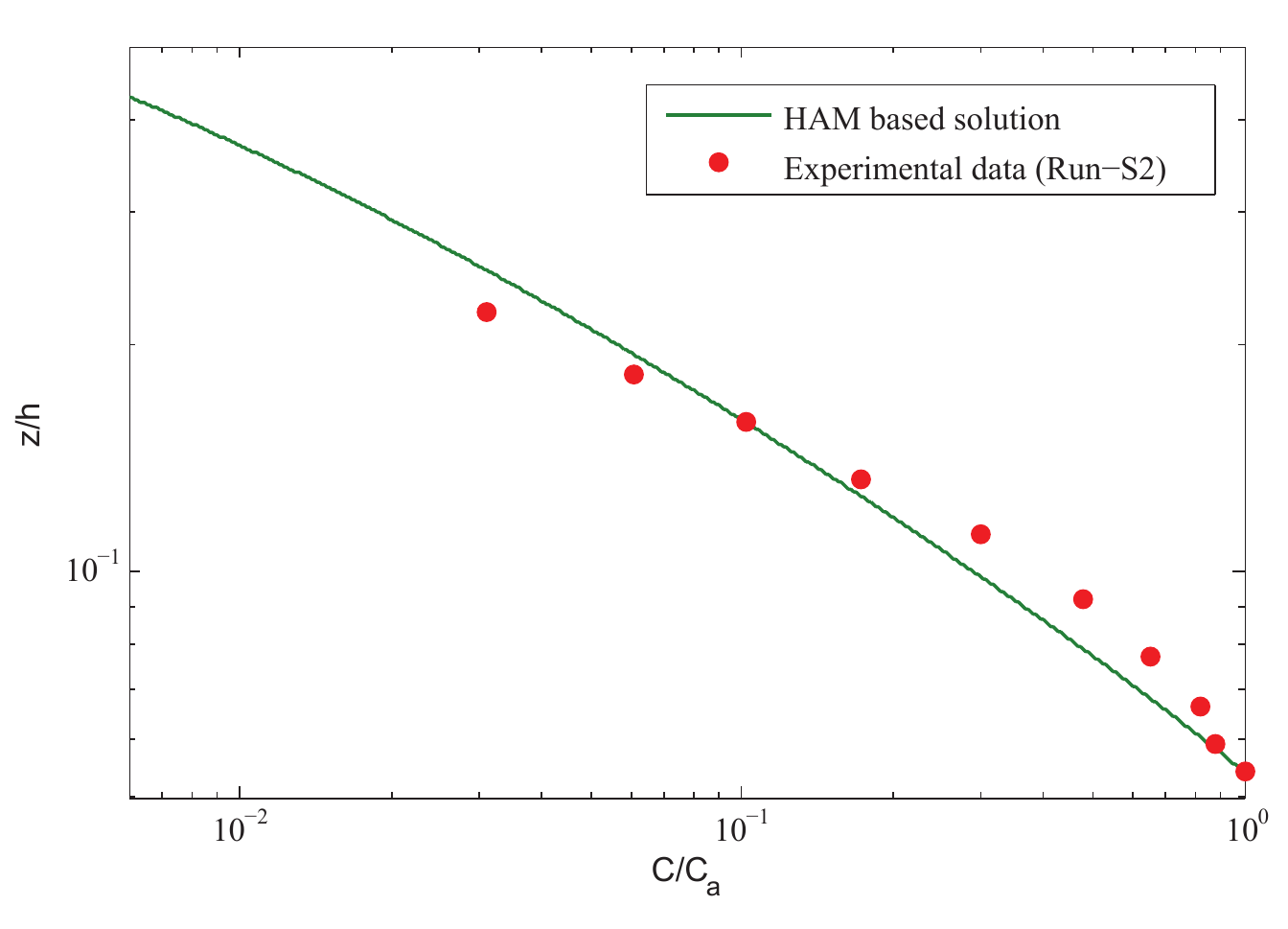}}
  \hfill
  \subfloat[]{\includegraphics[scale=0.33]{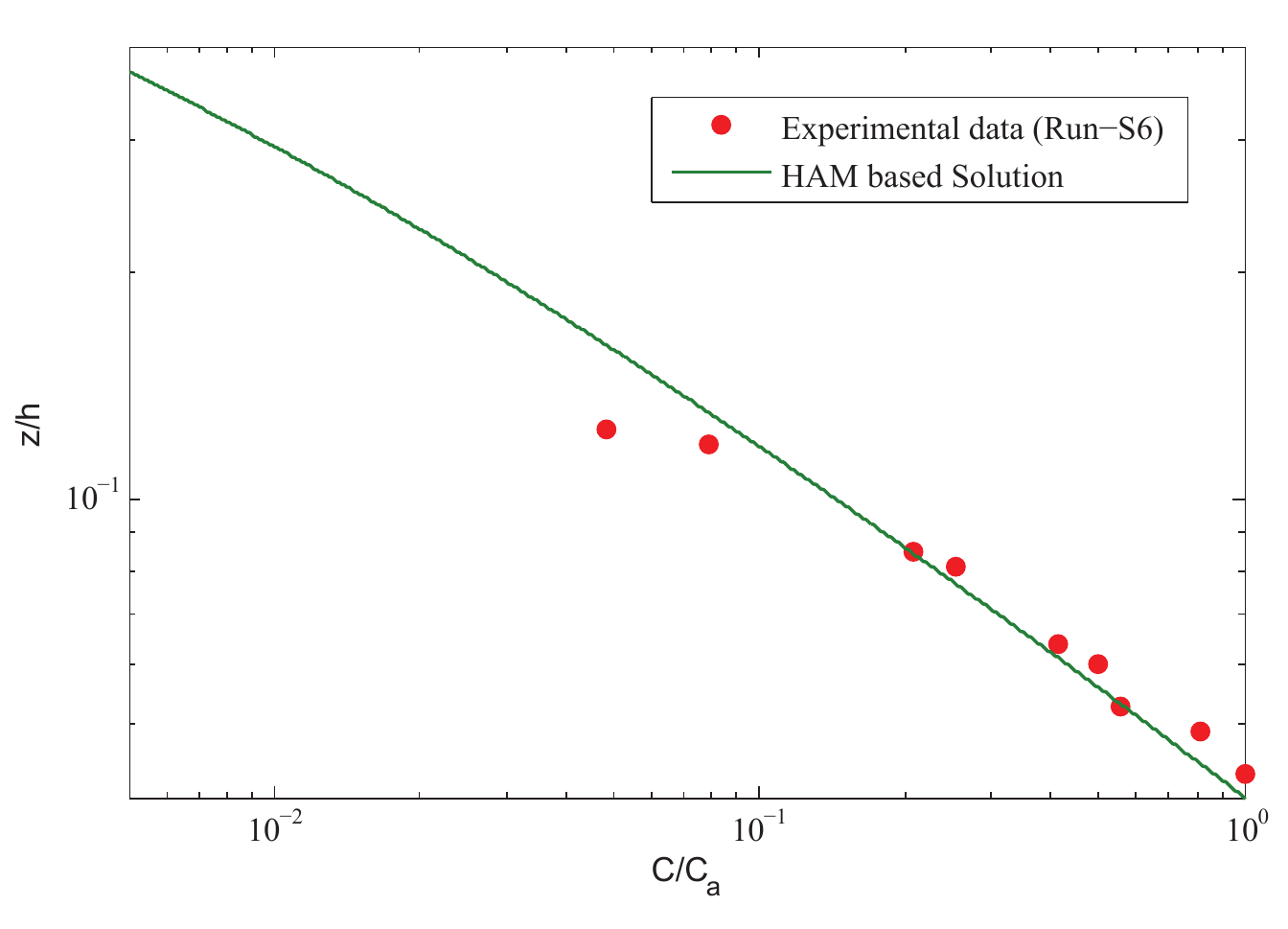}}
  \hfill
  \subfloat[]{\includegraphics[scale=0.33]{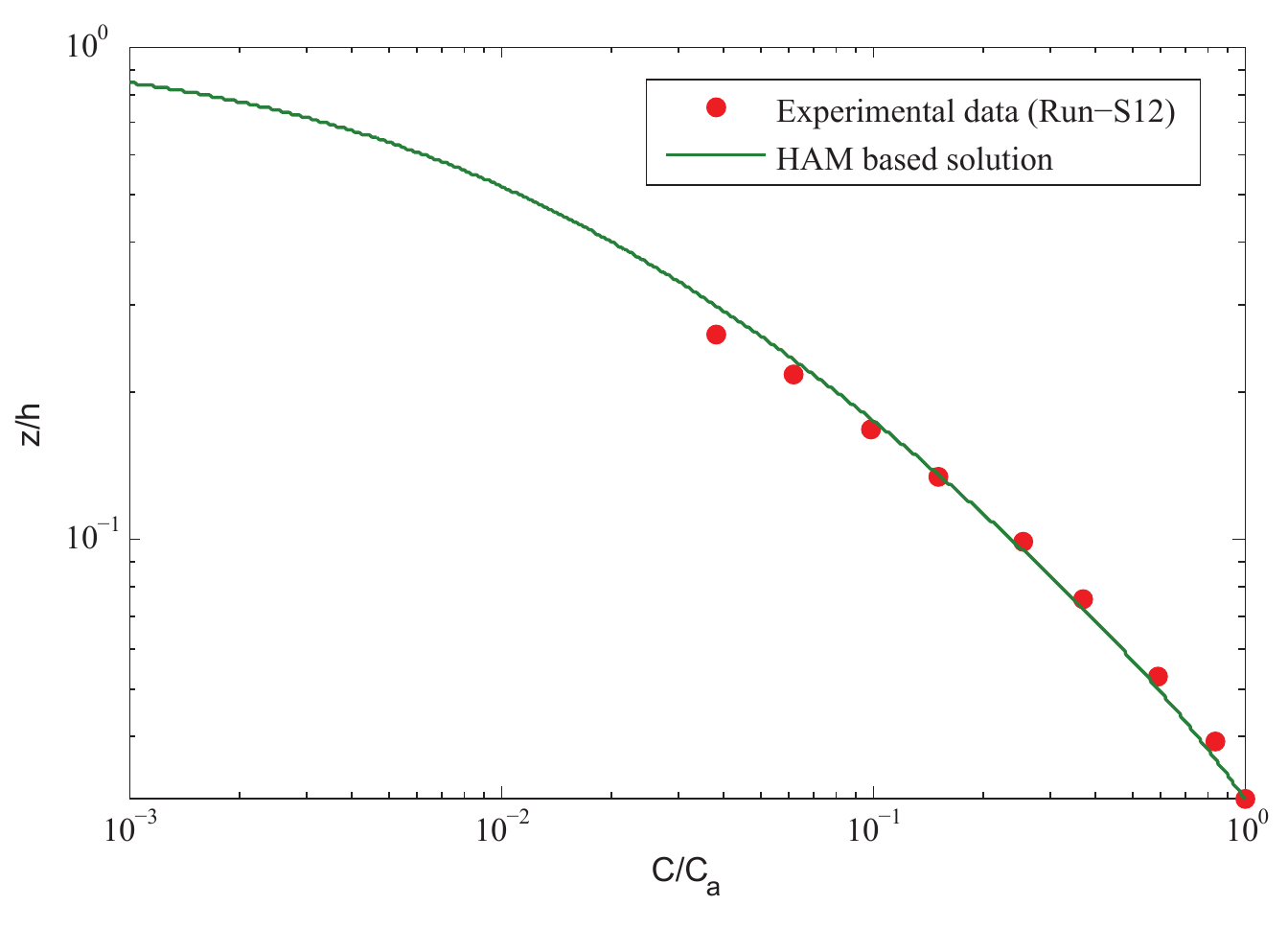}}
  \caption{Comparison between computed particle concentration profile and observed data of Einstein and Chien \cite{einstein1955effects}  (a) Run-S2, (b) Run-S6 and (c) Run-S12} \label{fig:Eins}
  \end{figure}

In this section, we compare the proposed model with experimental data available in literature. To that purpose, experiential data of Coleman \cite{coleman1986effects}, Lyn \cite{lyn1988similarity}, Vanoni \cite{vanoni1946transportation} and Einstein and Chien \cite{einstein1955effects} have been considered. Figs. \ref{fig:Col}, \ref{fig:lyn}, \ref{fig:Vanoni} and \ref{fig:Eins} compare the present model with selected runs from the data of Coleman \cite{coleman1986effects}, Lyn \cite{lyn1988similarity}, Vanoni \cite{vanoni1946transportation} and Einstein and Chien \cite{einstein1955effects}, respectively. It can be seen from the figures that the present model agrees well with observed values of concentration throughout the water depth.
\par
\par





\newpage
\section{Conclusions}
The following conclusions can be drawn from the present study:
\begin{enumerate}
\item The present work finds an explicit analytical solution to the modified Hunt \cite{hunt1954turbulent} equation for vertical distribution of suspended sediment concentration in an open channel flow. The diffusion coefficient is considered as the sum of sediment diffusion and shear-induced diffusion  coefficient, where the role of shear-induced diffusion coefficient lies in lifting the particles from bed-load layer to a small height above that layer. In addition, concentration-dependent settling velocity and non-unit ratio of sediment to turbulent diffusion coefficient is considered.
\item The homotopy analysis method is adopted for solving the highly non-linear governing equation analytically. The method is not confined to a particular class of differential equations and does not depend on small/large parameter present in the equation. Further, the convergence of the series solution can be tackled through convergence-control parameter.
\item It is found that the magnitude of shear-induced diffusion coefficient is more near the channel bed than in the main flow region and decays very fast after a short distance from the bed. On the other hand, effect of sediment diffusion coefficient is negligible near the bed and increases with the increase of distance from the bed. Also, it is observed that effect of shear-induced diffusion coefficient against concentration gradient is more compared to the shear-induced diffusion coefficient against shear gradient. Concentrations profiles have been plotted with and without the shear-induced diffusion coefficient and the concentration values are found to be relatively higher when shear-induced diffusion coefficient is present in the total diffusion coefficient.
\item The variation of shear-induced diffusion coefficient is found to be sensitive to the particle diameter and its magnitude becomes higher for larger particle. The effects of the ratio of sediment to turbulent diffusion coefficient and hindered settling mechanism on concentration profile are observed and interpreted physically.
\item Finally, the derived model is compared with numerical solution and it is seen that the analytical solution is close to numerical solution for higher order approximation. In addition, the solution obtained is compared with relevant sets of existing experimental data and the computed values are found to be in good agreement with the observed values of data.
    \end{enumerate}
\section*{\large{Appendix}}
\begin{thm}
Define an operator
\bea
\hat{\mathscr{D}}_{m}(\Phi)=\frac{1}{m!}\frac{\partial^{m}\Phi}{\partial q^{m}}
\eea
For a smooth function $f\in C^{\infty}(a,b)$ and a homotopy-Maclaurin series
\bea
\Phi=\sum_{k=0}^{\infty}\Phi_{k}q^{k}
\eea
it hold
\bea
\hat{\mathscr{D}}_{0}(\Phi)&=&f(\Phi)\nn\\
\hat{\mathscr{D}}_{m}(\Phi)&=&\mathlarger{\sum}_{k=0}^{m-1}(1-\frac{k}{m})\hat{\mathscr{D}}_{m-k}(\Phi)\frac{\partial}{\partial \Phi}{\mathscr{D}_{k}[f(\Phi)]}
\eea
and
\bea
\mathscr{D}_{m}[f(\Phi)]&=&\hat{\mathscr{D}}_{m}[f(\Phi)]\bigm|_{q=0}
\eea
\begin{proof}
It is obvious that $\hat{\mathscr{D}}_{0}(\Phi)=f(\Phi)$ holds. In case of $m\geq 1$, we have by Leibnitz's rule for derivative of product that
\bea
\hat{\mathscr{D}}_{m}[f(\Phi)]&=&\frac{1}{m!}\frac{\partial^{m}f(\Phi)}{\partial q^{m}}=\frac{1}{m!}\frac{\partial^{m-1}}{\partial q^{m-1}}\Bigg[\frac{\partial\Phi}{\partial q}\frac{\partial f(\Phi)}{\partial q}\Bigg]\nn\\
&=&\frac{1}{m!}\mathlarger{\sum}_{k=0}^{m-1}\frac{(m-1)!}{k!(m-1-k)!}\frac{\partial^{m-1-k}}{\partial q^{m-1-k}}\Bigg(\frac{\partial \Phi}{\partial q}\Bigg)\frac{\partial^{k}}{\partial q^{m}}\Bigg[\frac{\partial f(\Phi)}{\partial \Phi}\Bigg]\nn\\
&=&\mathlarger{\sum}_{k=0}^{m-1}\frac{m-k}{m}\Bigg[\frac{1}{(m-k)!}\frac{\partial^{m-k}\Phi}{\partial q^{m-k}}\Bigg]\Bigg\{\frac{1}{k!}\frac{\partial^{k}}{\partial q^{k}}\Bigg[\frac{\partial f(\Phi)}{\partial\Phi}\Bigg]\Bigg\}\nn\\
&=&\mathlarger{\sum}_{k=0}^{m-1}(1-\frac{k}{m})\hat{\mathscr{D}}_{m-k}(\Phi)\frac{\partial}{\partial \Phi}{\mathscr{D}_{k}[f(\Phi)]}
\eea
Since
\bea
\hat{\mathscr{D}}_{k}\Bigg[\frac{\partial f(\Phi)}{\partial\Phi}\Bigg]&=&\frac{\partial}{\partial \Phi}{\hat{\mathscr{D}}_{k}[f(\Phi)]}
\eea
\end{proof}
\end{thm}
\bibliographystyle{unsrt}

\end{document}